

\input amssym.def

\message{Cross-reference macros, B. Davies, version 5 May 1993.}

\catcode`@=11


\newif\if@xrf\@xrffalse   
\def\l@bel #1 #2 #3>>{\expandafter\gdef\csname @@#1#2\endcsname{#3}}
\immediate\newread\xrffile
\def\xrf@n#1#2{\expandafter\expandafter\expandafter
\csname immediate\endcsname\csname #1\endcsname\xrffile#2}
\def\xrf@@n{\if@xrf\relax\else%
  \expandafter\xrf@n{openin}{ = \jobname.xrf}\relax%
  \ifeof\xrffile%
    \message{ no file \jobname.xrf - run again for correct forward references
}%
  \else%
    \expandafter\xrf@n{closein}{}\relax%
    \setbox0=\hbox{\catcode`@=11 \input\jobname.xrf \catcode`@=12}%
  \fi\global\@xrftrue%
  \expandafter\expandafter\csname immediate\endcsname%
  \csname  newwrite\endcsname\xrffile%
  \expandafter\xrf@n{openout}{ = \jobname.xrf}\relax\fi}


\newcount\t@g

\def\order#1{%
  \expandafter\expandafter\csname newcount\endcsname
  \csname t@ghd#1\endcsname\csname t@ghd#1\endcsname=0

  \expandafter\def\csname #1\endcsname##1{\xrf@@n\csname n@#1\endcsname##1:>}

  \expandafter\def\csname n@#1\endcsname##1:##2>%
    {\def\n@xt{##1}\ifx\n@xt\empty%
     \expandafter\csname n@@#1\endcsname##1:##2:>
     \else\def\n@xt{##2}\ifx\n@xt\empty%
     \expandafter\csname n@@#1\endcsname\unp@ck##1 >:##2:>\else%
     \expandafter\csname n@@#1\endcsname\unp@ck##1 >:##2>\fi\fi}

  \expandafter\def\csname n@@#1\endcsname##1:##2:>%
    {\edef\t@g{\csname t@g#1\endcsname}\edef\t@@ghd{\csname t@ghd#1\endcsname}%
     \ifnum\t@@ghd=\t@ghd\else\global\t@@ghd=\number\t@ghd\global\t@g=0\fi%
     \ifunc@lled{@#1}{##1}\global\advance\t@g by 1%
       {\def\n@xt{##1}\ifx\n@xt\empty%
       \else\writ@new{#1}{##1}{\pret@g\t@ghead\number\t@g}\expandafter%
       \xdef\csname @#1##1\endcsname{\pret@g\t@ghead\number\t@g}\fi}%
       {\pret@g\t@ghead\number\t@g}%
     \else\def\n@xt{##1}%
       \w@rnmess#1,\n@xt>\csname @#1##1\endcsname%
     \fi##2}\ord@r{#1}}

\def\ord@r#1{%
  \expandafter\expandafter\csname newcount\endcsname
  \csname t@g#1\endcsname\csname t@g#1\endcsname=0

  \expandafter\def\csname ref#1\endcsname##1{%
     \expandafter\each@rg\csname #1c@te\endcsname{##1}}

  \expandafter\def\csname #1c@te\endcsname##1:##2,%
    {\def\n@xt{##2}\ifx\n@xt\empty%
     \csname #1cit@\endcsname##1:##2:,\else%
       \csname #1cit@\endcsname##1:##2,\fi}

  \expandafter\def\csname #1cit@\endcsname##1:##2:,%
    {\def\n@xt{\unp@ck##1 >}\ifunc@lled{@#1}{\n@xt}%
      {\expandafter\ifx\csname @@#1\n@xt\endcsname\relax%
       \und@fmess#1,\n@xt>>>\n@xt<<%
       \else\csname @@#1\n@xt\endcsname##2\fi}%
     \else\csname @#1\n@xt\endcsname##2%
     \fi}}


\def\sporder#1{%

  \expandafter\def\csname #1\endcsname##1{\xrf@@n\csname n@#1\endcsname##1:>}

  \expandafter\def\csname n@#1\endcsname##1:##2>%
    {\def\n@xt{##1}\ifx\n@xt\empty%
     \expandafter\csname n@@#1\endcsname##1:##2:>
     \else\def\n@xt{##2}\ifx\n@xt\empty%
     \expandafter\csname n@@#1\endcsname\unp@ck##1 >:##2:>\else%
     \expandafter\csname n@@#1\endcsname\unp@ck##1 >:##2>\fi\fi}

  \expandafter\def\csname n@@#1\endcsname##1:##2:>%
    {\edef\t@g{\csname t@g#1\endcsname}%
     \ifunc@lled{@#1}{##1}\global\advance\t@g by 1%
       {\def\n@xt{##1}\ifx\n@xt\empty%
       \else\writ@new{#1}{##1}{\number\t@g}\expandafter%
       \xdef\csname @#1##1\endcsname{\number\t@g}\fi}{\number\t@g}%
     \else\def\n@xt{##1}\w@rnmess#1,\n@xt>\csname @#1##1\endcsname%
     \fi##2}\ord@r{#1}}


\def\each@rg#1#2{{\let\thecsname=#1\expandafter\first@rg#2,\end,}}
\def\first@rg#1,{\callr@nge{#1}\apply@rg}
\def\apply@rg#1,{\ifx\end#1\let\n@xt=\relax%
\else,\callr@nge{#1}\let\n@xt=\apply@rg\fi\n@xt}

\def\callr@nge#1{\calldor@nge#1-\end-}
\def\callr@ngeat#1\end-{#1}
\def\calldor@nge#1-#2-{\ifx\end#2\thecsname#1:,%
  \else\thecsname#1:,\hbox{\rm--}\thecsname#2:,\callr@ngeat\fi}


\def\unp@ck#1 #2>{\unp@@k#1@> @>>}
\def\unp@@k#1 #2>>{\ifx#2@\@np@@k#1\else\@np@@k#1@> \unp@@k#2>>\fi}
\def\@np@@k#1#2#3>{\ifx#2@\@@np@@k#1>\else\@@np@@k#1>\@np@@k#2#3>\fi}
\def\@@np@@k#1>{\ifcat#1\alpha\expandafter\@@np@@@k\string#1>\else#1\fi}
\def\@@np@@@k#1#2>{@#2}


\def\writ@new#1#2#3{\xrf@@n\immediate\write\xrffile
  {\noexpand\l@bel #1 #2 {#3}>>}}


\def\ifunc@lled#1#2{\expandafter\ifx\csname #1#2\endcsname\relax}
\def\und@fmess#1#2,#3>{\ifx#1@%
  \message{ ** error - eqn label >>#3<< undefined - run again ** }\else
  \message{ ** error - #1#2 label >>#3<< undefined - run again ** }\fi}
\def\w@rnmess#1#2,#3>{\ifx#1@%
  \message{ Warning - duplicate eqn label >>#3<< }\else
  \message{ Warning - duplicate #1#2 label >>#3<< }\fi}


\def\t@ghead{}
\newcount\t@ghd\t@ghd=0
\def\taghe@d#1{\gdef\t@ghead{#1}\global\advance\t@ghd by 1}


\order{@qn}


\let\eqno@@=\eqno
\def\eqno(#1){\xrf@@n\eqno@@\hbox{{\rm(}$\@qn{#1}${\rm)}}}

\let\leqno@@=\leqno
\def\leqno(#1){\xrf@@n\leqno@@\hbox{{\rm(}$\@qn{#1}${\rm)}}}

\def\ref#1{\xrf@@n{{\rm(}$\ref@qn{#1}${\rm)}}}


\def\eq(#1){\xrf@@n\hfill\llap{\hbox{{\rm(}$\@qn{#1}${\rm)}}}}


\expandafter\csname newskip\endcsname\c@ntering
\c@ntering=0pt plus 1000pt minus 1000pt
\def\eqalignno#1{\xrf@@n\displ@y \tabskip=\c@ntering
  \halign to\displaywidth{\hfil$\displaystyle{##}$\tabskip=0pt
   &$\displaystyle{{}##}$\hfil\tabskip=\c@ntering
   &\llap{$\eqaln@##$}\tabskip=0pt\crcr
   #1\crcr}}
\def\leqalignno#1{\xrf@@n\displ@y \tabskip=\c@ntering
  \halign to\displaywidth{\hfil$\displaystyle{##}$\tabskip=0pt
   &$\displaystyle{{}##}$\hfil\tabskip=\c@ntering
    &\kern-\displaywidth\rlap{$\eqaln@##$}\tabskip\displaywidth\crcr
   #1\crcr}}
\def\eqaln@#1#2{\relax\ifcat#1(\expandafter\eqno@\else\fi#1#2}
\def\eqno@(#1){\xrf@@n\hbox{{\rm(}$\@qn{#1}${\rm)}}}


\def\n@@me#1#2>{#2}
\def\numberby#1{\xrf@@n
  \ifx\s@ction\undefined\else
  \expandafter\let\csname\s@@ve\endcsname=\s@ction\fi
  \ifx\subs@ction\undefined\else
  \expandafter\let\csname\subs@@ve\endcsname=\subs@ction\fi
  \numb@rby#1,>#1>}
\def\numb@rby#1,#2>#3>{\def\n@xt{#1}\ifx\n@xt\empty\taghe@d{}\else
  \def\n@xt{#2}\ifx\n@xt\empty\n@by#3>\else\n@@by#3>\fi\fi}
\def\n@by#1>{\ifx\s@cno\undefined\expandafter\expandafter
  \csname newcount\endcsname\csname s@cno\endcsname
  \csname s@cno\endcsname=0\else\s@cno=0\fi
  \xdef\s@@ve{\expandafter\n@@me\string#1>}
  \let\s@ction=#1\def#1{\global\advance\s@cno by 1
  \taghe@d{\number\s@cno.}\s@ction}}
\def\n@@by#1,#2>{\ifx\s@cno\undefined\expandafter\expandafter
  \csname newcount\endcsname\csname s@cno\endcsname
  \csname s@cno\endcsname=0\else\s@cno=0\fi
  \ifx\subs@cst\undefined\expandafter\expandafter
  \csname newcount\endcsname\csname subs@cst\endcsname
  \csname subs@cst\endcsname=0\else\subs@cst=0\fi
  \ifx\subs@cno\undefined\expandafter\expandafter
  \csname newcount\endcsname\csname subs@cno\endcsname
  \csname subs@cno\endcsname=0\else\subs@cno=0\fi
  \xdef\s@@ve{\expandafter\n@@me\string#1>}
  \let\s@ction=#1\def#1{\global\advance\s@cno by 1
  \global\subs@cno=\subs@cst
  \taghe@d{\number\s@cno.}\s@ction}
  \xdef\subs@@ve{\expandafter\n@@me\string#2>}
  \let\subs@ction=#2\def#2{\global\advance\subs@cno by 1
  \taghe@d{\number\s@cno.\number\subs@cno.}\subs@ction}}


\def\numberfrom#1{\ifx\s@cno\undefined\else\n@mberfrom#1,>\fi}
\def\n@mberfrom#1,#2>{\def\n@xt{#2}%
  \ifx\n@xt\empty\n@@f#1>\else\n@@@f#1,#2>\fi}
\def\n@@f#1>{\s@cno=#1\advance\s@cno by -1}
\def\n@@@f#1,#2,>{\s@cno=#1\advance\s@cno by -1%
  \subs@cst=#2\advance\subs@cst by -1}


\def\pret@g{}
\def\prefixby#1{\gdef\pret@g{#1}}



\newcount\r@fcount\r@fcount=0
\newcount\r@fcurr
\newcount\r@fmin
\newcount\r@fmax
\newcount\r@fone
\newcount\r@ftwo
\newif\ifc@te\c@tefalse
\newif\ifr@feat

\def\cite#1{{\rm[}\def\s@p{}\r@fmin=\r@fcount\r@fmax=0%
   \refn@te#1>>\r@fcurr=\r@fmin\advance\r@fcurr by-1\refc@te{\rm]}}


\def\refn@te#1>>{\refn@@te#1,>>}

\def\refn@@te#1,#2>>{\r@fnote{\str@pbl#1 >>}%
   \def\n@xt{#2}\ifx\n@xt\empty\else\refn@@te#2>>\fi}

\def\r@fnote#1%
  {\ifunc@lled{r@f}{#1}%
     \global\advance\r@fcount by 1\r@fmax=\r@fcount\r@fcurr=\r@fcount%
     \expandafter\xdef\csname r@f#1\endcsname{\number\r@fcount}%
     \expandafter\gdef\csname r@ftext\number\r@fcount\endcsname%
     {\message{ Reference #1 to be supplied }%
     Reference $#1$ to be supplied\par}%
   \else%
     \expandafter\r@fcurr=\csname r@f#1\endcsname\relax%
     \ifnum\r@fmin<\r@fcurr\else\r@fmin=\r@fcurr\advance\r@fmin by-1\fi%
     \ifnum\r@fmax<\r@fcurr\r@fmax=\r@fcurr\fi%
   \fi\expandafter\expandafter\def\csname r@fn\number\r@fcurr\endcsname{Y}}

\def\ifunc@lled#1#2{\expandafter\ifx\csname #1#2\endcsname\relax}

\def\str@pbl#1 #2>>{\str@@pbl#1#2 >>}
\def\str@@pbl#1 #2>>{\str@@@pbl#1#2 >>}
\def\str@@@pbl#1 #2>>{\str@@@@pbl#1#2 >>}
\def\str@@@@pbl#1 #2>>{#1}


\def\refc@te{\r@featfalse\def\s@ve{}%
  {\loop\ifnum\r@fcurr<\r@fmax\advance\r@fcurr by 1\c@tefalse%
   \expandafter\refc@@te\number\r@fcurr>>%
   \ifc@te\expandafter\refe@t\number\r@fcurr>>\fi\repeat\s@ve}}

\def\refc@@te#1>>{\ifnum#1=0\edef\n@xt{}\else
   \edef\n@xt{\csname r@fn#1\endcsname}\fi%
   \expandafter\xdef\csname r@fn#1\endcsname{}%
   \ifx\n@xt\empty\else\relax\c@tetrue\fi}

\def\refe@t#1>>{\ifr@feat\ifnum\r@fone=\r@ftwo\res@cond#1>>%
   \else\reth@rd#1>>\fi\else\r@feattrue\ref@rst#1>>\fi}

\def\ref@rst#1>>{\r@feattrue\r@fone=#1\r@ftwo=#1%
   \s@p\expandafter\relax\number\r@fone}%

\def\res@cond#1>>{\advance\r@ftwo by 1\def\n@xt{#1}%
   \expandafter\ifnum\n@xt=\number\r@ftwo%
   \edef\s@ve{,\expandafter\relax\number\r@ftwo}%
   \else\def\s@p{,}\ref@rst#1>>\fi}%

\def\reth@rd#1>>{\advance\r@ftwo by 1\def\n@xt{#1}%
   \expandafter\ifnum\n@xt=\number\r@ftwo%
   \edef\s@ve{--\expandafter\relax\number\r@ftwo}\def\s@p{,}\else%
   \def\s@p{,}\s@ve\def\s@ve{}\ref@rst#1>>\fi}%


\def\refis#1 #2\par{\ifunc@lled{r@f}{#1}\else
   \expandafter\gdef\csname r@ftext\csname
r@f#1\endcsname\endcsname{#2\par}\fi}


\def\listreferences{%
\ifx\numrefjl\undefined
  \csname newdimen\endcsname\@ldskip\@ldskip=\parskip
  \csname newdimen\endcsname\@ldindent\@ldindent=\parindent
  \def\numr@f##1>>##2>>{\leftskip=40pt\parskip=0pt\parindent=-10pt%
    \hskip-30pt\hbox to 30pt{\rm[##1]\hss}##2\par\vskip\@ldskip%
    \parskip=\@ldskip\parindent=\@ldindent\leftskip=0pt}
  {\global\r@fcurr=0%
   \loop\ifnum\r@fcurr<\r@fcount\global\advance\r@fcurr by 1%
   \numr@f\number\r@fcurr>>\csname r@ftext\number\r@fcurr\endcsname>>\repeat}
\else
  {\global\r@fcurr=0%
   \loop\ifnum\r@fcurr<\r@fcount\global\advance\r@fcurr by 1%
   \def\refnum{\number\r@fcurr}\csname r@ftext\number\r@fcurr\endcsname\repeat}
\fi}


\def\beginsection#1\par{\vskip0pt plus.2\vsize\penalty-250
  \vskip0pt plus -.2\vsize\bigskip\vskip\parskip
  \leftline{\bf#1}\nobreak\smallskip\noindent}

\catcode`@=12


%
\catcode`@=11
\newif\ifnews@ct\news@ctfalse
\parskip=2pt plus 2pt
\font\twelvebf=cmbx10 scaled 1200
\font\twelveit=cmti10 scaled 1200
\newcount\secno\secno=0
\def\section#1#2\par#3\par
  {\vskip2cm\penalty-250\vskip-2cm\bigskip
  \if#1*\noindent{\bf#2}\else
  \advance\secno by 1\subsecno=0\news@cttrue
  \noindent\hbox to \parindent{\bf\number\secno.\hfil}{\bf#1}\fi
  \par\vskip-\parskip\medskip#3\news@ctfalse\par}
\newcount\subsecno\subsecno=0
\def\subsection#1
  {\advance\subsecno by 1
  \ifnews@ct\else\smallskip\fi
  \noindent\number\secno.\number\subsecno\hskip1ex{\sl #1.}\hskip1ex}
%
%

\def\id{{\rm id\/}}
\def\tr{{\rm tr\/}}
\def\wt{{\rm wt\/}}

\def\Hom{{\rm Hom\/}}

\def\Phit{\tilde{\Phi}}
\def\mapright#1{\mathop{\longrightarrow}\limits^{#1}}
\def\XXZ{{\rm X\kern-.1pt X\kern-.3pt Z}}
\def\CTM{{\rm C\kern-.3pt T\kern-.3pt M}}
\def\BCTM{{\bf C\kern-.3pt T\kern-.3pt M}}
\def\HX{H_{{\rm XX\kern-.3pt Z}}}
\def\KX{K_{{\rm XX\kern-.3pt Z}}}
\def\KC{K_{{\rm CT\kern-.3pt M}}}
\def\VO{{V\kern-.5pt O}}
\def\ABF{{\rm A\kern-.3pt B\kern-.4pt F}}
\def\slt{\gs\gl_2}
\def\slth{\widehat{\slt}}
\def\U{U_q\bigl(\slt\bigr)}
\def\Up{U'_q\bigl(\slth\bigr)}
\def\Uq{U_q\bigl(\slth\bigr)}
\def\weight#1#2#3#4#5{\hgrid=6pt\vgrid=16pt
  \gridcommdiag{&&&\hskip-1.6pt{}^#4\cr\lower3pt\hbox{${}^#1$}&&
  \raise2.2pt\hbox{$\drawline(-1,0){20pt}$}
  \hskip-10pt\lower7.8pt\hbox{$\drawline(0,1){20pt}$}&&
  &&\lower3pt\hbox{${}^#3$}\cr&&&\hskip-1.6pt{}_#2\cr}
  \lower1pt\hbox{$#5$}}
\def\triangle{\hbox{
  \lower8pt\hbox{$\drawline(6,5){30pt}$}\hskip-30pt
  \raise8pt\hbox{$\drawline(6,-5){30pt}$}\hskip-10pt
  \lower20pt\hbox{$\drawline(0,1){40pt}$}}}
\def\revtriangle{\hbox{
  \lower17pt\hbox{$\drawline(6,5){30pt}$}\hskip-30pt
  \raise17pt\hbox{$\drawline(6,-5){30pt}$}\hskip-20pt
  \lower20pt\hbox{$\drawline(0,1){40pt}$}}}
\def\vertex#1#2#3#4{\hgrid=6pt\vgrid=16pt\sarrowlength=20pt
  \gridcommdiag{&&&\hskip-1.6pt\raise2pt\hbox{$#2$}\cr
  \hskip-10pt#3&&\sline(1,0)\hskip-.5\sarrowlength\sline(0,1)&&&&#1\cr
  &&&\hskip-1.6pt#4\cr}}
\def\dcross{\hbox{
  \lower10pt\hbox{$\drawline(1,1){20pt}$}\hskip-20pt
  \raise10pt\hbox{$\drawline(1,-1){20pt}$}}}
\def\para{\hbox{
  \lower10pt\hbox{$\drawline(1,0){20pt}$}\hskip-20pt
  \raise10pt\hbox{$\drawline(1,0){20pt}$}}}
\def\hcross{\sline(1,0)\hskip-.5\sarrowlength\sline(0,1)}
\def\alab#1{{\hskip-6pt\alpha_#1}}
\def\blab#1{{\hskip-6pt\beta_#1}}
\def\clab#1{{\hskip-6pt\gamma_#1}}

%
%
\ifx\Bbb\undefined
  \message{Black board bold font not found in macros.tex}
  \def\BC{{\bf C}}
  
  \def\BQ{{\bf Q}}
  \def\BZ{{\bf Z}}
\else
  \def\BC{{\Bbb C}}
  
  \def\BQ{{\Bbb Q}}
  \def\BZ{{\Bbb Z}}
\fi
\ifx\goth\undefined
  \message{Gothic font not found in macros.tex}
  \def\gs{{\sl s}}
  \def\gl{{\sl l}}
  \def\gH{{\sl H}}
\else
  \def\gs{{\goth s}}
  \def\gl{{\goth l}}
  \def\gH{{\goth H}}
\fi
%
%
\catcode`@=12
%
%
\catcode`@=11
\newif\ifs@p
\def\jnlitem#1#2#3#4%
  {#1\def\l@st{#1}\ifx\l@st\empty\s@pfalse\else\s@ptrue\fi%
   \def\l@st{#2}\ifx\l@st\empty\else%
   \ifs@p, \fi{\frenchspacing\sl#2}\s@ptrue\fi%
   \def\l@st{#3}\ifx\l@st\empty\else\ifs@p, \fi{\bf#3}\s@ptrue\fi%
   \def\l@st{#4}\ifx\l@st\empty\else\ifs@p, \fi#4\s@ptrue\fi%
   \ifs@p.\fi\par}
\def\bkitem#1#2#3%
  {#1\def\l@st{#1}\ifx\l@st\empty\s@pfalse\else\s@ptrue\fi%
   \def\l@st{#2}\ifx\l@st\empty\else%
   \ifs@p, \fi{\frenchspacing\sl#2}\s@ptrue\fi%
   \def\l@st{#3}\ifx\l@st\empty\else\ifs@p, \fi#3\s@ptrue\fi%
   \ifs@p.\fi\par}
\catcode`@=12
%
%

\def\APNY{Ann. Phys., NY}

\def\CMP{Comm. Math. Phys.}
\def\Duke{Duke Math. J.}

\def\IJMPA{Int. J. Mod. Phys. A}

\def\JMP{J. Math. Phys.}
\def\JPA{J. Phys. A: Math. Gen.}
\def\JPC{J. Phys. C: Solid State Phys.}
\def\JSP{J. Stat. Phys.}
\def\LMP{Lett. Math. Phys.}

\def\NPB{Nucl. Phys. B}

\def\PA{Physica A}
\def\PD{Physica D}

\def\PLA{Phys. Lett. A}
\def\PLB{Phys. Lett. B}
\def\PR{Phys. Rev.}

\def\PRL{Phys. Rev. Lett.}

\def\PTP{Prog. Theor. Phys.}
\def\RMP{Rev. Mod. Phys.}

\edef\thinlines{\the\catcode`@ }%
\catcode`@ = 11
\let\@oldatcatcode = \thinlines
\edef\@oldandcatcode{\the\catcode`& }%
\catcode`& = 11
\def\&whilenoop#1{}%
\def\&whiledim#1\do #2{\ifdim #1\relax#2\&iwhiledim{#1\relax#2}\fi}%
\def\&iwhiledim#1{\ifdim #1\let\&nextwhile=\&iwhiledim
        \else\let\&nextwhile=\&whilenoop\fi\&nextwhile{#1}}%
\newif\if&negarg
\newdimen\&wholewidth
\newdimen\&halfwidth
\font\tenln=line10
\def\thinlines{\let\&linefnt\tenln \let\&circlefnt\tencirc
  \&wholewidth\fontdimen8\tenln \&halfwidth .5\&wholewidth}%
\def\thicklines{\let\&linefnt\tenlnw \let\&circlefnt\tencircw
  \&wholewidth\fontdimen8\tenlnw \&halfwidth .5\&wholewidth}%
\def\drawline(#1,#2)#3{\&xarg #1\relax \&yarg #2\relax \&linelen=#3\relax
  \ifnum\&xarg =0 \&vline \else \ifnum\&yarg =0 \&hline \else \&sline\fi\fi}%
\def\&sline{\leavevmode
  \ifnum\&xarg< 0 \&negargtrue \&xarg -\&xarg \&yyarg -\&yarg
  \else \&negargfalse \&yyarg \&yarg \fi
  \ifnum \&yyarg >0 \&tempcnta\&yyarg \else \&tempcnta -\&yyarg \fi
  \ifnum\&tempcnta>6 \&badlinearg \&yyarg0 \fi
  \ifnum\&xarg>6 \&badlinearg \&xarg1 \fi
  \setbox\&linechar\hbox{\&linefnt\&getlinechar(\&xarg,\&yyarg)}%
  \ifnum \&yyarg >0 \let\&upordown\raise \&clnht\z@
  \else\let\&upordown\lower \&clnht \ht\&linechar\fi
  \&clnwd=\wd\&linechar
  \&whiledim \&clnwd <\&linelen \do {%
    \&upordown\&clnht\copy\&linechar
    \advance\&clnht \ht\&linechar
    \advance\&clnwd \wd\&linechar
  }%
  \advance\&clnht -\ht\&linechar
  \advance\&clnwd -\wd\&linechar
  \&tempdima\&linelen\advance\&tempdima -\&clnwd
  \&tempdimb\&tempdima\advance\&tempdimb -\wd\&linechar
  \hskip\&tempdimb \multiply\&tempdima \@m
  \&tempcnta \&tempdima \&tempdima \wd\&linechar \divide\&tempcnta \&tempdima
  \&tempdima \ht\&linechar \multiply\&tempdima \&tempcnta
  \divide\&tempdima \@m
  \advance\&clnht \&tempdima
  \ifdim \&linelen <\wd\&linechar \hskip \wd\&linechar
  \else\&upordown\&clnht\copy\&linechar\fi}%
\def\&hline{\vrule height \&halfwidth depth \&halfwidth width \&linelen}%
\def\&getlinechar(#1,#2){\&tempcnta#1\relax\multiply\&tempcnta 8
  \advance\&tempcnta -9 \ifnum #2>0 \advance\&tempcnta #2\relax\else
  \advance\&tempcnta -#2\relax\advance\&tempcnta 64 \fi
  \char\&tempcnta}%
\def\drawvector(#1,#2)#3{\&xarg #1\relax \&yarg #2\relax
  \&tempcnta \ifnum\&xarg<0 -\&xarg\else\&xarg\fi
  \ifnum\&tempcnta<5\relax \&linelen=#3\relax
    \ifnum\&xarg =0 \&vvector \else \ifnum\&yarg =0 \&hvector
    \else \&svector\fi\fi\else\&badlinearg\fi}%
\def\&hvector{\ifnum\&xarg<0 \rlap{\&linefnt\&getlarrow(1,0)}\fi \&hline
  \ifnum\&xarg>0 \llap{\&linefnt\&getrarrow(1,0)}\fi}%
\def\&vvector{\ifnum \&yarg <0 \&downvector \else \&upvector \fi}%
\def\&svector{\&sline
  \&tempcnta\&yarg \ifnum\&tempcnta <0 \&tempcnta=-\&tempcnta\fi
  \ifnum\&tempcnta <5
    \if&negarg\ifnum\&yarg>0                   
      \llap{\lower\ht\&linechar\hbox to\&linelen{\&linefnt
        \&getlarrow(\&xarg,\&yyarg)\hss}}\else 
      \llap{\hbox to\&linelen{\&linefnt\&getlarrow(\&xarg,\&yyarg)\hss}}\fi
    \else\ifnum\&yarg>0                        
      \&tempdima\&linelen \multiply\&tempdima\&yarg
      \divide\&tempdima\&xarg \advance\&tempdima-\ht\&linechar
      \raise\&tempdima\llap{\&linefnt\&getrarrow(\&xarg,\&yyarg)}\else
      \&tempdima\&linelen \multiply\&tempdima-\&yarg 
      \divide\&tempdima\&xarg
      \lower\&tempdima\llap{\&linefnt\&getrarrow(\&xarg,\&yyarg)}\fi\fi
  \else\&badlinearg\fi}%
\def\&getlarrow(#1,#2){\ifnum #2 =\z@ \&tempcnta='33\else
\&tempcnta=#1\relax\multiply\&tempcnta \sixt@@n \advance\&tempcnta
-9 \&tempcntb=#2\relax\multiply\&tempcntb \tw@
\ifnum \&tempcntb >0 \advance\&tempcnta \&tempcntb\relax
\else\advance\&tempcnta -\&tempcntb\advance\&tempcnta 64
\fi\fi\char\&tempcnta}%
\def\&getrarrow(#1,#2){\&tempcntb=#2\relax
\ifnum\&tempcntb < 0 \&tempcntb=-\&tempcntb\relax\fi
\ifcase \&tempcntb\relax \&tempcnta='55 \or
\ifnum #1<3 \&tempcnta=#1\relax\multiply\&tempcnta
24 \advance\&tempcnta -6 \else \ifnum #1=3 \&tempcnta=49
\else\&tempcnta=58 \fi\fi\or
\ifnum #1<3 \&tempcnta=#1\relax\multiply\&tempcnta
24 \advance\&tempcnta -3 \else \&tempcnta=51\fi\or
\&tempcnta=#1\relax\multiply\&tempcnta
\sixt@@n \advance\&tempcnta -\tw@ \else
\&tempcnta=#1\relax\multiply\&tempcnta
\sixt@@n \advance\&tempcnta 7 \fi\ifnum #2<0 \advance\&tempcnta 64 \fi
\char\&tempcnta}%
\def\&vline{\ifnum \&yarg <0 \&downline \else \&upline\fi}%
\def\&upline{\hbox to \z@{\hskip -\&halfwidth \vrule width \&wholewidth
   height \&linelen depth \z@\hss}}%
\def\&downline{\hbox to \z@{\hskip -\&halfwidth \vrule width \&wholewidth
   height \z@ depth \&linelen \hss}}%
\def\&upvector{\&upline\setbox\&tempboxa\hbox{\&linefnt\char'66}\raise
     \&linelen \hbox to\z@{\lower \ht\&tempboxa\box\&tempboxa\hss}}%
\def\&downvector{\&downline\lower \&linelen
      \hbox to \z@{\&linefnt\char'77\hss}}%
\def\&badlinearg{\errmessage{Bad \string\arrow\space argument.}}%
\thinlines
\countdef\&xarg     0
\countdef\&yarg     2
\countdef\&yyarg    4
\countdef\&tempcnta 6
\countdef\&tempcntb 8
\dimendef\&linelen  0
\dimendef\&clnwd    2
\dimendef\&clnht    4
\dimendef\&tempdima 6
\dimendef\&tempdimb 8
\chardef\@arrbox    0
\chardef\&linechar  2
\chardef\&tempboxa  2           
\let\lft^%
\let\rt_
\newif\if@pslope 
\def\@findslope(#1,#2){\ifnum#1>0
  \ifnum#2>0 \@pslopetrue \else\@pslopefalse\fi \else
  \ifnum#2>0 \@pslopefalse \else\@pslopetrue\fi\fi}%
\def\generalsmap(#1,#2){\getm@rphposn(#1,#2)\plnmorph\futurelet\next\addm@rph}%
\def\sline(#1,#2){\setbox\@arrbox=\hbox{\drawline(#1,#2){\sarrowlength}}%
  \@findslope(#1,#2)\d@@blearrfalse\generalsmap(#1,#2)}%
\def\arrow(#1,#2){\setbox\@arrbox=\hbox{\drawvector(#1,#2){\sarrowlength}}%
  \@findslope(#1,#2)\d@@blearrfalse\generalsmap(#1,#2)}%
\newif\ifd@@blearr
\def\bisline(#1,#2){\@findslope(#1,#2)%
  \if@pslope \let\@upordown\raise \else \let\@upordown\lower\fi
  \getch@nnel(#1,#2)\setbox\@arrbox=\hbox{\@upordown\@vchannel
    \rlap{\drawline(#1,#2){\sarrowlength}}%
      \hskip\@hchannel\hbox{\drawline(#1,#2){\sarrowlength}}}%
  \d@@blearrtrue\generalsmap(#1,#2)}%
\def\biarrow(#1,#2){\@findslope(#1,#2)%
  \if@pslope \let\@upordown\raise \else \let\@upordown\lower\fi
  \getch@nnel(#1,#2)\setbox\@arrbox=\hbox{\@upordown\@vchannel
    \rlap{\drawvector(#1,#2){\sarrowlength}}%
      \hskip\@hchannel\hbox{\drawvector(#1,#2){\sarrowlength}}}%
  \d@@blearrtrue\generalsmap(#1,#2)}%
\def\adjarrow(#1,#2){\@findslope(#1,#2)%
  \if@pslope \let\@upordown\raise \else \let\@upordown\lower\fi
  \getch@nnel(#1,#2)\setbox\@arrbox=\hbox{\@upordown\@vchannel
    \rlap{\drawvector(#1,#2){\sarrowlength}}%
      \hskip\@hchannel\hbox{\drawvector(-#1,-#2){\sarrowlength}}}%
  \d@@blearrtrue\generalsmap(#1,#2)}%
\newif\ifrtm@rph
\def\@shiftmorph#1{\hbox{\setbox0=\hbox{$\scriptstyle#1$}%
  \setbox1=\hbox{\hskip\@hm@rphshift\raise\@vm@rphshift\copy0}%
  \wd1=\wd0 \ht1=\ht0 \dp1=\dp0 \box1}}%
\def\@hm@rphshift{\ifrtm@rph
  \ifdim\hmorphposnrt=\z@\hmorphposn\else\hmorphposnrt\fi \else
  \ifdim\hmorphposnlft=\z@\hmorphposn\else\hmorphposnlft\fi \fi}%
\def\@vm@rphshift{\ifrtm@rph
  \ifdim\vmorphposnrt=\z@\vmorphposn\else\vmorphposnrt\fi \else
  \ifdim\vmorphposnlft=\z@\vmorphposn\else\vmorphposnlft\fi \fi}%
\def\addm@rph{\ifx\next\lft\let\temp=\lftmorph\else
  \ifx\next\rt\let\temp=\rtmorph\else\let\temp\relax\fi\fi \temp}%
\def\plnmorph{\dimen1\wd\@arrbox \ifdim\dimen1<\z@ \dimen1-\dimen1\fi
  \vcenter{\box\@arrbox}}%
\def\lftmorph\lft#1{\rtm@rphfalse \setbox0=\@shiftmorph{#1}%
  \if@pslope \let\@upordown\raise \else \let\@upordown\lower\fi
  \llap{\@upordown\@vmorphdflt\hbox to\dimen1{\hss 
    \llap{\box0}\hss}\hskip\@hmorphdflt}\futurelet\next\addm@rph}%
\def\rtmorph\rt#1{\rtm@rphtrue \setbox0=\@shiftmorph{#1}%
  \if@pslope \let\@upordown\lower \else \let\@upordown\raise\fi
  \llap{\@upordown\@vmorphdflt\hbox to\dimen1{\hss
    \rlap{\box0}\hss}\hskip-\@hmorphdflt}\futurelet\next\addm@rph}%
\def\getm@rphposn(#1,#2){\ifd@@blearr \dimen@\morphdist \advance\dimen@ by
  .5\channelwidth \@getshift(#1,#2){\@hmorphdflt}{\@vmorphdflt}{\dimen@}\else
  \@getshift(#1,#2){\@hmorphdflt}{\@vmorphdflt}{\morphdist}\fi}%
\def\getch@nnel(#1,#2){\ifdim\hchannel=\z@ \ifdim\vchannel=\z@
    \@getshift(#1,#2){\@hchannel}{\@vchannel}{\channelwidth}%
    \else \@hchannel\hchannel \@vchannel\vchannel \fi
  \else \@hchannel\hchannel \@vchannel\vchannel \fi}%
\def\@getshift(#1,#2)#3#4#5{\dimen@ #5\relax
  \&xarg #1\relax \&yarg #2\relax
  \ifnum\&xarg<0 \&xarg -\&xarg \fi
  \ifnum\&yarg<0 \&yarg -\&yarg \fi
  \ifnum\&xarg<\&yarg \&negargtrue \&yyarg\&xarg \&xarg\&yarg \&yarg\&yyarg\fi
  \ifcase\&xarg \or  
    \ifcase\&yarg    
      \dimen@i \z@ \dimen@ii \dimen@ \or 
      \dimen@i .7071\dimen@ \dimen@ii .7071\dimen@ \fi \or
    \ifcase\&yarg    
      \or 
      \dimen@i .4472\dimen@ \dimen@ii .8944\dimen@ \fi \or
    \ifcase\&yarg    
      \or 
      \dimen@i .3162\dimen@ \dimen@ii .9486\dimen@ \or
      \dimen@i .5547\dimen@ \dimen@ii .8321\dimen@ \fi \or
    \ifcase\&yarg    
      \or 
      \dimen@i .2425\dimen@ \dimen@ii .9701\dimen@ \or\or
      \dimen@i .6\dimen@ \dimen@ii .8\dimen@ \fi \or
    \ifcase\&yarg    
      \or 
      \dimen@i .1961\dimen@ \dimen@ii .9801\dimen@ \or
      \dimen@i .3714\dimen@ \dimen@ii .9284\dimen@ \or
      \dimen@i .5144\dimen@ \dimen@ii .8575\dimen@ \or
      \dimen@i .6247\dimen@ \dimen@ii .7801\dimen@ \fi \or
    \ifcase\&yarg    
      \or 
      \dimen@i .1645\dimen@ \dimen@ii .9864\dimen@ \or\or\or\or
      \dimen@i .6402\dimen@ \dimen@ii .7682\dimen@ \fi \fi
  \if&negarg \&tempdima\dimen@i \dimen@i\dimen@ii \dimen@ii\&tempdima\fi
  #3\dimen@i\relax #4\dimen@ii\relax }%
\catcode`\&=4  
\def\generalhmap{\futurelet\next\@generalhmap}%
\def\@generalhmap{\ifx\next^ \let\temp\generalhm@rph\else
  \ifx\next_ \let\temp\generalhm@rph\else \let\temp\m@kehmap\fi\fi \temp}%
\def\generalhm@rph#1#2{\ifx#1^
    \toks@=\expandafter{\the\toks@#1{\rtm@rphtrue\@shiftmorph{#2}}}\else
    \toks@=\expandafter{\the\toks@#1{\rtm@rphfalse\@shiftmorph{#2}}}\fi
  \generalhmap}%
\def\m@kehmap{\mathrel{\smash{\the\toks@}}}%
\def\mapright{\toks@={\mathop{\vcenter{\smash{\drawrightarrow}}}\limits}%
  \generalhmap}%
\def\mapleft{\toks@={\mathop{\vcenter{\smash{\drawleftarrow}}}\limits}%
  \generalhmap}%
\def\bimapright{\toks@={\mathop{\vcenter{\smash{\drawbirightarrow}}}\limits}%
  \generalhmap}%
\def\bimapleft{\toks@={\mathop{\vcenter{\smash{\drawbileftarrow}}}\limits}%
  \generalhmap}%
\def\adjmapright{\toks@={\mathop{\vcenter{\smash{\drawadjrightarrow}}}\limits}%
  \generalhmap}%
\def\adjmapleft{\toks@={\mathop{\vcenter{\smash{\drawadjleftarrow}}}\limits}%
  \generalhmap}%
\def\hline{\toks@={\mathop{\vcenter{\smash{\drawhline}}}\limits}%
  \generalhmap}%
\def\bihline{\toks@={\mathop{\vcenter{\smash{\drawbihline}}}\limits}%
  \generalhmap}%
\def\drawrightarrow{\hbox{\drawvector(1,0){\harrowlength}}}%
\def\drawleftarrow{\hbox{\drawvector(-1,0){\harrowlength}}}%
\def\drawbirightarrow{\hbox{\raise.5\channelwidth
  \hbox{\drawvector(1,0){\harrowlength}}\lower.5\channelwidth
  \llap{\drawvector(1,0){\harrowlength}}}}%
\def\drawbileftarrow{\hbox{\raise.5\channelwidth
  \hbox{\drawvector(-1,0){\harrowlength}}\lower.5\channelwidth
  \llap{\drawvector(-1,0){\harrowlength}}}}%
\def\drawadjrightarrow{\hbox{\raise.5\channelwidth
  \hbox{\drawvector(-1,0){\harrowlength}}\lower.5\channelwidth
  \llap{\drawvector(1,0){\harrowlength}}}}%
\def\drawadjleftarrow{\hbox{\raise.5\channelwidth
  \hbox{\drawvector(1,0){\harrowlength}}\lower.5\channelwidth
  \llap{\drawvector(-1,0){\harrowlength}}}}%
\def\drawhline{\hbox{\drawline(1,0){\harrowlength}}}%
\def\drawbihline{\hbox{\raise.5\channelwidth
  \hbox{\drawline(1,0){\harrowlength}}\lower.5\channelwidth
  \llap{\drawline(1,0){\harrowlength}}}}%
\def\generalvmap{\futurelet\next\@generalvmap}%
\def\@generalvmap{\ifx\next\lft \let\temp\generalvm@rph\else
  \ifx\next\rt \let\temp\generalvm@rph\else \let\temp\m@kevmap\fi\fi \temp}%
\toksdef\toks@@=1
\def\generalvm@rph#1#2{\ifx#1\rt 
    \toks@=\expandafter{\the\toks@
      \rlap{$\vcenter{\rtm@rphtrue\@shiftmorph{#2}}$}}\else 
    \toks@@={\llap{$\vcenter{\rtm@rphfalse\@shiftmorph{#2}}$}}%
    \toks@=\expandafter\expandafter\expandafter{\expandafter\the\expandafter
      \toks@@ \the\toks@}\fi \generalvmap}%
\def\m@kevmap{\the\toks@}%
\def\mapdown{\toks@={\vcenter{\drawdownarrow}}\generalvmap}%
\def\mapup{\toks@={\vcenter{\drawuparrow}}\generalvmap}%
\def\bimapdown{\toks@={\vcenter{\drawbidownarrow}}\generalvmap}%
\def\bimapup{\toks@={\vcenter{\drawbiuparrow}}\generalvmap}%
\def\adjmapdown{\toks@={\vcenter{\drawadjdownarrow}}\generalvmap}%
\def\adjmapup{\toks@={\vcenter{\drawadjuparrow}}\generalvmap}%
\def\vline{\toks@={\vcenter{\drawvline}}\generalvmap}%
\def\bivline{\toks@={\vcenter{\drawbivline}}\generalvmap}%
\def\drawdownarrow{\hbox to5pt{\hss\drawvector(0,-1){\varrowlength}\hss}}%
\def\drawuparrow{\hbox to5pt{\hss\drawvector(0,1){\varrowlength}\hss}}%
\def\drawbidownarrow{\hbox to5pt{\hss\hbox{\drawvector(0,-1){\varrowlength}}%
  \hskip\channelwidth\hbox{\drawvector(0,-1){\varrowlength}}\hss}}%
\def\drawbiuparrow{\hbox to5pt{\hss\hbox{\drawvector(0,1){\varrowlength}}%
  \hskip\channelwidth\hbox{\drawvector(0,1){\varrowlength}}\hss}}%
\def\drawadjdownarrow{\hbox to5pt{\hss\hbox{\drawvector(0,-1){\varrowlength}}%
  \hskip\channelwidth\lower\varrowlength
  \hbox{\drawvector(0,1){\varrowlength}}\hss}}%
\def\drawadjuparrow{\hbox to5pt{\hss\hbox{\drawvector(0,1){\varrowlength}}%
  \hskip\channelwidth\raise\varrowlength
  \hbox{\drawvector(0,-1){\varrowlength}}\hss}}%
\def\drawvline{\hbox to5pt{\hss\drawline(0,1){\varrowlength}\hss}}%
\def\drawbivline{\hbox to5pt{\hss\hbox{\drawline(0,1){\varrowlength}}%
  \hskip\channelwidth\hbox{\drawline(0,1){\varrowlength}}\hss}}%
\def\commdiag#1{\null\,
  \vcenter{\commdiagbaselines
  \m@th\ialign{\hfil$##$\hfil&&\hfil$\mkern4mu ##$\hfil\crcr
      \mathstrut\crcr\noalign{\kern-\baselineskip}
      #1\crcr\mathstrut\crcr\noalign{\kern-\baselineskip}}}\,}%
\def\commdiagbaselines{\baselineskip15pt \lineskip3pt \lineskiplimit3pt }%
\def\gridcommdiag#1{\null\,
  \vcenter{\offinterlineskip
  \m@th\ialign{&\vbox to\vgrid{\vss
    \hbox to\hgrid{\hss\smash{$##$}\hss}}\crcr
      \mathstrut\crcr\noalign{\kern-\vgrid}
      #1\crcr\mathstrut\crcr\noalign{\kern-.5\vgrid}}}\,}%
\newdimen\harrowlength \harrowlength=60pt
\newdimen\varrowlength \varrowlength=.618\harrowlength
\newdimen\sarrowlength \sarrowlength=\harrowlength
\newdimen\hmorphposn \hmorphposn=\z@
\newdimen\vmorphposn \vmorphposn=\z@
\newdimen\morphdist  \morphdist=4pt
\dimendef\@hmorphdflt 0       
\dimendef\@vmorphdflt 2       
\newdimen\hmorphposnrt  \hmorphposnrt=\z@
\newdimen\hmorphposnlft \hmorphposnlft=\z@
\newdimen\vmorphposnrt  \vmorphposnrt=\z@
\newdimen\vmorphposnlft \vmorphposnlft=\z@

\newdimen\hgrid \hgrid=15pt
\newdimen\vgrid \vgrid=15pt
\newdimen\hchannel  \hchannel=0pt
\newdimen\vchannel  \vchannel=0pt
\newdimen\channelwidth \channelwidth=3pt
\dimendef\@hchannel 0         
\dimendef\@vchannel 2         

\magnification 1200
\hsize=15.6truecm
\vsize=23truecm
\newdimen\loffset\loffset=.3truecm
\newdimen\roffset\roffset=.3truecm
\voffset=1truecm
\parskip=2pt
\nopagenumbers
\pageno=0
\headline={\ifnum\pageno=0{}\global\hoffset=\loffset\else \hdline\fi}
\def\hdline{\ifodd\pageno\rightheadline \else\leftheadline\fi}
\def\rightheadline{\tenrm\hfil{\it \rhead}\hfil\folio\global\hoffset=\loffset}
\def\leftheadline{\tenrm\folio\hfil{\it \lhead}\hfil\global\hoffset=\roffset}
%
%
\def\ltitle{Infinite dimensional symmetry \cr
           of corner transfer matrices \cr}
\def\stitle{Infinite dimensional symmetry of corner transfer matrices}

\def\lauthor{Brian Davies}

\def\laddress{Department of Mathematics, \hfil\break
School of Mathematical Sciences, \hfil\break
Australian National University, \hfil\break
Canberra, ACT 0200, Australia }

\def\SMS{SMS-118-93}
\def\CMA{CMA-MR55-93}

\def\labstract{
We review some of the recent developments in two dimensional
statistical mechanics in which corner transfer matrices
provide the vital link between the physical system and the
representation theory of quantum affine algebras.
This opens many new possibilities, because the eigenstates may
be described using the properties of $q$-vertex operators.}
%
%
\def\lhead{\lauthor}
\def\rhead{\stitle}
\vglue 1cm
\centerline{\twelvebf\vbox{\halign{\hfil # \quad\hfil\cr\ltitle\crcr}}}
\vglue 1cm
\centerline{\twelveit\vbox{\halign{\hfil # \quad\hfil\cr\lauthor\crcr}}}
\vglue 1cm
\centerline{\hskip2.2cm\SMS\hfill\CMA\hskip3cm}
\vglue 2cm
\noindent\laddress
\vglue 1cm
\noindent December 1993
\vglue 1.5cm
\noindent 1991 {\it Mathematics subject classification}:
primary 82B13; secondary 82B20.
\vglue 1.5cm
{\narrower
\font\smr=cmr10 scaled 937
\font\smb=cmbx10 scaled 937
\noindent{\smb Abstract.}\hskip1ex\smr\labstract\par}
\vfill\eject
%
%

\section{Introduction}

Transfer matrices have been a central object for the study of
two-dimensional statistical mechanics systems for more than a half century.
Originally the row transfer matrix was defined: it
encapsulates the statistical information (via the Boltzmann weights)
pertaining to the contribution from a single row of a regular lattice.
Row transfer matrices are extremely complicated; for an Ising or six-vertex
model on a square lattice of width $L$ the row transfer matrix has
dimension $2^L$. Onsager's solution of the Ising model \cite{Ons44} used
representation theory of finite-dimensional Lie algebras to transform the
generators of the row transfer matrix to a tractable form.
The next several major advances --- for example the solution of the
one-dimensional Bose gas \cite{LL63}, the six \cite{L67} and eight
\cite{Bax71} vertex models --- brought a totally different approach to the
forefront, the Bethe Ansatz.
But developments in the last decade have restored the r\^ole of
representation theory.
The quantum inverse scattering method \cite{Fad81,Th81}, an algebraic
implementation of the Bethe Ansatz and Yang-Baxter equations,
gave birth to the theory of quantum groups \cite{J85,D87}.
Representation theory of infinite dimensional Lie
algebras became paramount in conformal field theory \cite{BPZ84,FC91}.

An independent development was the invention of the corner
transfer matrix (\CTM) \cite{Bax76,Bax81}.
It is the partition function for a whole corner of a regular
lattice and might therefore be expected to be even less tractable than
the row transfer matrix.
Surprisingly, this is not so: \CTM s have an extremely simple eigenvalue
spectrum (integer powers of a single parameter).
\CTM s proved to be an effective method for the evaluation of one-point
functions \cite{Bax81,Bax80,ABF84}.
In \cite{ABF84} configuration sums appeared which were identified as
characters of Virasoro algebras \cite{Car84,FQS84}.
These mysterious connections led to the discovery of hierarchies of
solvable lattice models in two dimensions \cite{JMO88a} and of many
beautiful connections with infinite dimensional Lie algebras
\cite{DJKMO87,JMO88b,DJKMO89}.
Furthermore, the multiplicities of the \CTM\ eigenvalues
are equal to the weight-space multiplicities
of irreducible highest weight representations of certain affine Lie
algebras \cite{DJKMO89}, and the actual configurations
used to label the \CTM\ eigenvectors also label the crystal base
vectors of the corresponding representations
\cite{Ka90,Ka91,MM90,JMMO91}.
These were vital clues for finding the links with representation theory.
But the crucial point was to identify the appropriate algebras and their
representations, and the r\^ole of the row transfer matrices and \CTM s in
these algebras \cite{FM92,DFJMN92}.

In one of the original papers on \CTM s \cite{Bax76}, Baxter observed that
there is no Ising-like reduction in the six and eight vertex models.  He
wrote

{\narrower\noindent\it A rather ambitious hope is that by examining the
\CTM s we may stumble on such a group, that the solution of the models may
thereby be simplified ...\par}

\noindent  In \cite{DFJMN92} was given a new scheme for solving the
six-vertex model, in the antiferromagnetic regime, using the newly
discovered quantum affine symmetry of the \CTM.
The approach of that paper has been extended to higher spin chains
\cite{IIJMNT92}, to the higher rank case \cite{DO93} and to the \ABF\
models of Andrews, Baxter and Forrester \cite{JMO92}.
The scheme has already been used in some of those models to give general
expressions for the $n$-point correlation functions
\cite{JMMN92}: it also provides a means to derive $q$-difference
equations for them \cite{JMMN92,FJMMN93}.
The promise of this new approach is self-evident.
Many hard-won results of Bethe Ansatz calculations have been
recovered, important new results have  been obtained, and there is
much research in progress, since the calculation of correlation functions
is a long-standing problem.

The six-vertex model is related to the quantum affine algebra
$\Uq$ of Drinfeld and Jimbo \cite{D87,J89}, since its Boltzmann weights
are the simplest example of an $R$-matrix.
It is a parad\-igm for the application of the quantum affine symmetry to
the representation of \CTM s.
Briefly, the magic of the six-vertex \CTM\ arises from the property that
its eigenvectors provide level-$1$ highest weight
modules of the quantum affine algebra $\Uq$, while the \CTM\
acts as a derivation operator in the algebra.
Similar statements may be made in the other cases (higher spin, higher
rank).
Stated more simply, the Chevalley generators of the quantum affine algebra
are raising and lowering operators when acting on the eigenvectors of the
\CTM, and the importance of this cannot be over-stated.
It is precisely the kind of property which makes the Ising model so
tractable, although the fermion algebra is much simpler than the quantum
affine algebra with which we deal here.

The purpose of this article is to give an introductory account which is as
elementary as possible.
It is based on talks which have been given by the author.
For simplicity we consider in detail only the six-vertex model on
which the original ideas were developed.
In addition, the focus is on the r\^ole of \CTM s; consequently we
omit any discussion of the construction and properties of
creation/annihilation operators for the \XXZ\ Hamiltonian, although this
was a major goal in \cite{DFJMN92}.
There is a growing literature concerned with the extension of the methods
to various statistical mechanics systems and with the use of
$q$-difference equations which are an essential ingredient in the theory
of $q$-vertex operators.
The six-vertex model uses the level-$1$ modules of the rank-$1$ quantum
affine algebra $\Up$.
Higher-spin and higher-rank cases are considered in
\cite{IIJMNT92,DO93,K93}.
The Boltzmann weights of all these models form the $R$-matrices which
intertwine tensor products of finite-dimensional representations.
The relation is more complicated for the \ABF\ models: the
Boltzmann weights are the connection coefficients which intertwine vertex
operators, themselves intertwiners \cite{FR92}.
The procedure for the \ABF\ models \cite {JMO92} is somewhat
analogous to the Goddard-Kent-Olive coset construction of the discrete
minimal series of Virasoro algebras \cite{GKO85}.
For the many details of all these developments the
reader should consult the cited literature.

\section{Transfer matrices}

\subsection{Definition of the model}
The six-vertex model is defined on a regular lattice, each vertex of which is
at the intersection of two edges, which carry the physical spin variables
$\pm$.
The energy $E_{\rm c}$ of a configuration of the system derives is the sum
of the energy $E_{\rm v}$ assigned to each vertex.
{}From this additivity, the Boltzmann weight $\exp(-E_{\rm c}/kT)$ of a
configuration is the product  $\prod_{\rm v}\exp(-E_{\rm v}/kT)$
of vertex weights
$$
\weight\alpha\beta\gamma\delta{\qquad=\qquad
W(\alpha,\beta,\gamma,\delta;u_{\alpha\gamma}-u_{\beta\delta}).}
$$
The six-vertex model is ``$Z$-invariant'' in the sense of Baxter
\cite{Bax78}.
For this reason we attach directions and rapidity variables
$u_{\alpha\gamma}$,  $u_{\beta\delta}$ to the lines of the lattice, so that
$W(\alpha,\beta,\gamma,\delta;u)$ has as
its rapidity their difference.
The convention is that negating a direction negates the corresponding
$u_{\alpha\gamma}$ or $u_{\beta\delta}$.

Only six of the  Boltzmann weights are non-zero.
In the notation of \cite{Baxbk} they are $a=\rho\sinh(\lambda -u)$,
$b=\rho\sinh u$, $c=\rho\sinh\lambda$, where $\rho$ is a normalisation factor
and $0\le u\le\lambda$ in the physical regime.
In anticipation of the connection with the quantum affine algebra, we
introduce new variables by
$$
q=-\exp(-\lambda),\qquad
\zeta=\exp(u),\qquad
\Delta={q+q^{-1}\over2}.
$$
Also, we normalise the weights by the partition function per
site, $\kappa(\zeta)$:
$$
\weight++++{\quad=\quad}
\weight ----{\quad=\quad
\displaystyle{1\over\kappa(\zeta)},\,
\phantom{\displaystyle{1-q^2\over1-q^2\zeta^2}\,\zeta}}
$$
$$
\weight+-+-{\quad=\quad}
\weight-+-+{\quad=\quad
\displaystyle{1\over\kappa(\zeta)}\,
\displaystyle{1-\zeta^2\over1-q^2\zeta^2}\,q,}
$$
$$
\weight+--+{\quad=\quad}
\weight-++-{\quad=\quad
\displaystyle{1\over\kappa(\zeta)}\,
\displaystyle{1-q^2\over1-q^2\zeta^2}\,\zeta.}
$$

\subsection{Properties of vertex weights}
The most important property of the weights is that they satisfy the Yang-Baxter
equations, which have the well-known pictorial representation
$$
\triangle\hskip1.6cm=\hskip1cm\revtriangle\qquad
$$
This is the origin of $Z$-invariance.
Summation is implied over all internal lines and this sum is invariant under
rearrangement of the lines, provided only that all vertices have the same
value of $\lambda$ and the boundary arrangement is fixed.
Because of the locality of the Yang-Baxter equations, this holds for
arbitrarily large systems.

Two other important properties are unitarity
$$
\hgrid=10pt\vgrid=10pt
\gridcommdiag{
\cr\dcross&&\para&&\dcross&&&
\lower2.5pt\hbox{$=$}&&&\para&&\para\cr\cr}
$$
and crossing symmetry
$$
\weight{\alpha}{\beta}{\gamma}{\delta}{~~(\zeta)} \quad=\quad
\weight{{-\delta}}{\alpha}{{-\beta}}{\gamma}{~~(-q^{-1}\zeta^{-1})}
$$

The partition function per site may be obtained from these by the standard
inversion trick \cite{Baxbk}.
We sketch the argument.
Unitarity and crossing symmetry imply the relations
$$
\kappa(\zeta)\,\kappa(\zeta^{-1})=1,\qquad
\kappa(-q^{-1}\zeta^{-1})=
{1-q^2\zeta^2\over q(1-\zeta^2)}\,\kappa(\zeta).
$$
This gives a $q$-difference equation and the unique solution for a function
analytic in the annulus $q^2\le|\zeta|\le q^{-1}$, and normalised to
$\kappa(1)=1$, is
$$
\kappa(\zeta)=\zeta
{(\zeta^2q^4;q^4)_\infty(\zeta^{-2}q^2;q^4)_\infty\over
(\zeta^{-2}q^4;q^4)_\infty(\zeta^2q^2;q^4)_\infty},
$$
where
$$
(z;p)_\infty=\prod_{n=0}^\infty(1-zp^n).
$$

\subsection{Corner transfer matrix}
\CTM s are defined as the partition function of a quadrant of the lattice.
The configurations of the edges of the quadrant label the rows and columns of
the \CTM\ whilst the outside boundary, which is part of the boundary of the
total system, is fixed so as to select the required ground state.
There is a preferred diagonal direction ({\it SE--NW}), which
is along the boundary of $A(u)$, and two ground states labelled
$i=0,1$, related by a lattice shift of one site.

\CTM s are defined in the infinite lattice limit in a normalised
form \cite{Baxbk}.
A small part of the construction of two \CTM s, $A(u)$, $B(v)$ looks like:
$$
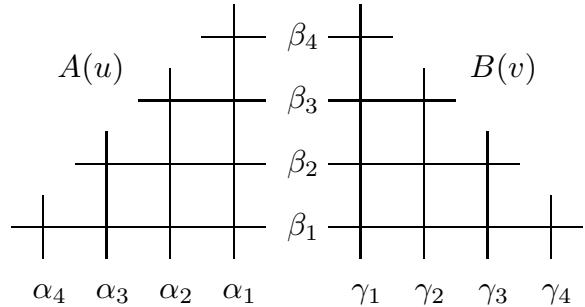
\hgrid=10pt\vgrid=10pt\sarrowlength=20pt
\gridcommdiag{\cr
  &&&&&&&&\hcross&&&\blab4&\hcross\cr
  &&&&A(u)&&&&&&&&&&&&&B(v)\cr
  &&&&&&\hcross&&\hcross&&&\blab3&\hcross&&\hcross\cr\cr
  &&&&\hcross&&\hcross&&\hcross&&&\blab2&\hcross&&\hcross&&\hcross\cr\cr
  &&\hcross&&\hcross&&\hcross&&\hcross&&&\blab1
  &\hcross&&\hcross&&\hcross&&\hcross\cr\cr
  &&&\alab4&&\alab3&&\alab2&&\alab1&&
  &&\clab1&&\clab2&&\clab3&&\clab4\cr}
$$
We  consistently use the notation for matrix elements of an operator
$L:V\rightarrow W$ that $Lv_\alpha=L_\alpha^\beta w_\beta$: the upper index
labels the row.

Recall Baxter's arguments from which the properties of the \CTM s stem.
He considers the product of two \CTM s $A(u)$, $B(v)$.
The result is the partition function of an infinite half plane and
its value is a matrix element of an arbitrarily high power of the row
transfer matrix.
This  depends only on the maximal eigenvector, which in turn is a function
only of the rapidity difference $u-v$ (Yang-Baxter equations).
This implies $B(v)A(u)=X(u-v)$.
The group property $A(u)A(v)=A(u+v)$ is the most important conclusion to be
drawn.
It follows from the crossing symmetry of the Boltzmann weights and spin
reversal.
The result is
$$
A(\zeta)=\zeta^{\textstyle{\KX}},\qquad
B(\zeta)=(-q\zeta)^{\textstyle{-\KX}}R,
$$
where
$$
\KX={-2\over\sqrt{\Delta^2-1}}
\sum_{n=1}^\infty n (\sigma^x_{n+1}\sigma^x_n +\sigma^y_{n+1}\sigma^y_n
+\Delta\sigma^z_{n+1}\sigma^z_n).
$$
and $R$ is the operation of spin reversal at every site.

\section{The quantum affine algebra}

\subsection{Notation}
In this section we define our notations for $\Uq$ and review those
properties needed for our discussion of the six-vertex model.
$\Up$ is generated by $e_i$, $f_i$, $t_i=q^{h_{i}}$,
$(i=0,1)$, which satisfy the defining relations \cite{J92}
$$
\eqalign{
t_i e_j &= q^{A_{ij}} e_j t_i,\qquad
t_i f_j = q^{-A_{ij}} f_j t_i, \cr
t_i t_j &=t_j t_i, \hskip39pt
[e_i,f_j]=\delta_{ij}{t_i-t_i^{-1}\over q-q^{-1}}, \cr}
$$
$$
\eqalign{
e_i^3 e_j - [3]e_i^2e_je_i + [3] e_ie_je_i^2 - e_je_i^3 &= 0, \cr
f_i^3 f_j - [3]f_i^2f_jf_i + [3]f_if_jf_i^2 - f_jf_i^3 &= 0, \cr}
\quad(i\neq j).
$$
Here $[n] = (q^n-q^{-n})/(q-q^{-1})$, and $A_{ij}$ is the generalised Cartan
matrix for the affine Lie algebra $\slth$
$$
A_{ij}=\pmatrix{2&-2\cr-2&2\cr}.
$$
To obtain the full quantum affine algebra $\Uq$ one must add the
generator $q^d$. $d$ is the derivation, and it satisfies
$$
[d,e_i]=\delta_{i,0}e_i, \qquad [d,f_i]=-\delta_{i,0}f_i.
\eqno(CD)
$$
The co-multiplication for the Hopf algebra structure is defined by
$$
\eqalign{
\Delta(e_i)&=e_i\otimes\id+t_i\otimes e_i, \cr
\Delta(f_i)&=f_i\otimes t_i^{-1}+\id\otimes f_i, \cr
\Delta(t_i)&=t_i\otimes t_i. \cr}
$$
and the formula for the antipode is
$$
a(e_i)=-t_i^{-1}e_i, \qquad
a(f_i)=-f_it_i, \qquad
a(t_i)=t_i^{-1}.
$$

\subsection{The $R$-matrix}
The simplest spin half $\U$ module $V$ over the field $\BQ(q)$
has basis vectors $v_{+}=v_0$, $v_{-}=v_1$, and is equipped with the actions
$$
\eqalign{e_1v_{+} &= 0,\cr e_1v_{-} &= v_{+},\cr}\qquad
\eqalign{f_1v_{+} &= v_{-},\cr f_1v_{-} &= 0,\cr}\qquad
\eqalign{t_1v_{+} &= qv_{+},\cr t_1v_{-} &= q^{-1}v_{-}.\cr}
$$
{}From this one may construct a $\Uq$ module by introducing the formal
variable $z$.
The basis vectors of this infinite-dimensional module are
$\{z^n\otimes v_\pm\}$ but we simply write $\{z^n v_\pm\}$.
That is, one equips $V_z=\BQ[z,z^{-1}]\otimes V$ with the $\Uq$
module action
$$
\eqalign{ e_0&\mapsto zf_1, \cr e_1&\mapsto e_1, \cr} \qquad
\eqalign{ f_0&\mapsto z^{-1}e_1, \cr f_1&\mapsto f_1, \cr} \qquad
\eqalign{ t_0&\mapsto t_1^{-1}, \cr t_1&\mapsto t_1, \cr} \qquad
d\mapsto z{d\over dz}.
$$
$V_z$ is also a $\Up$ module: for this one simply regards $z$ as a
fixed non-zero complex number and drops the grading operator $d$ (evaluation
representation).

Define the operator $R(z)$ by
$$
\eqalign{
R(z)&v_\pm\otimes v_\pm=v_\pm\otimes v_\pm,\cr
R(z)&v_{+}\otimes v_{-}=
 {1-z\over1-q^2z}q\,v_{+}\otimes v_{-}
 +{1-q^2\over1-q^2z}z\,v_{-}\otimes v_{+},\cr
R(z)&v_{-}\otimes v_{+}=
 {1-q^2\over1-q^2z}\,v_{+}\otimes v_{-}
 +{1-z\over1-q^2z}q\,v_{-}\otimes v_{+}.\cr}
$$
Further, let $P$ be the transposition operator:
$P\cdot(v_1\otimes v_2)=(v_2\otimes v_1)$.
Then
$\check{R}(z_1/z_2)=PR(z_1/z_2) :
V_{z_1}\otimes V_{z_2}\rightarrow V_{z_2}\otimes V_{z_1}$
is an intertwiner of $\Up$ modules.
Its entries also provide the Boltzmann weights for the six-vertex model
with a different normalisation
and a change of gauge.
Writing $z=\zeta^2$:
$$
W(i,j,k,l;\zeta)=\kappa(z)^{-1}g_{ik}^{-1}\check{R}_{ij}^{kl}(z),
$$
with $g_{ik}=1$, $i=k$, and  $g_{\pm\mp}=\zeta^{\pm1}$.
The gauge factor is the well-known term which makes finite spin chains
$\U$-invariant.
It simply adds the total spin to $\KX$ and changes its r\^ole as a grading
operator.
Since it does not affect the physics, we henceforth use the Boltzmann weights
$$
\overline{W}(i,j,k,l;z)=g_{ik}W(i,j,k,l;\zeta).
$$
This is consistent with the cited works
\cite{DFJMN92,IIJMNT92,JMO92,DO93,K93}.

\subsection{Dual modules}
The dual module construction is intimately connected with the crossing
symmetry and is required for representing local operators.
It goes as follows.
Given a linear space $V$ its dual $V^*$ is the set of linear maps
$V^*:V\rightarrow\BC$ by the action
$v~\mathop{\longmapsto}\limits^{\textstyle{u^*}}
{}~\langle u^*,v\rangle$.
Choose the dual basis by the canonical pairing
$\langle v^*_j,v_k\rangle=\delta_{jk}$.
$V^*_z$ becomes a dual module under the action
$$
\langle xu^*,v\rangle=\langle u^*,\phi(x) v\rangle,
\qquad\forall\,x\in\Uq,
$$
where $\phi$ is any anti-homomorphism of the algebra
($\phi(xy)=\phi(y)\phi(x)$).
Using the antipode for $\phi$, the action of $\Uq$ on
$V^*_z$ is
$$
\eqalign{&e_1v^*_{+} = -q^{-1}v^*_{-},\cr
&e_1v^*_{-} = 0,\cr
&e_0 \mapsto (zq^{-2})f_1, \cr} \qquad
\eqalign{&f_1v^*_{+} = 0,\cr
&f_1v^*_{-} = -qv^*_{-},\cr
&f_0 \mapsto (zq^{-2})^{-1}e_1, \cr} \qquad
\eqalign{&t_1v^*_{+} = q^{-1}v^*_{+},\cr
&t_1v^*_{-} = qv^*_{-},\cr
&t_0\mapsto t_1^{-1}. \cr}
$$
This gives the isomorphism
$$
C: V_{z}^*\mathop{\longrightarrow}\limits^\sim
V_{zq^{-2}},\qquad
v^*_{+}\mapsto v_{-},\qquad
v^*_{-}\mapsto -qv_{+}.
$$

The operator $R(z)$ was defined to act on
$V_{z_1}\otimes V_{z_2}$ so that $PR(z)$ is an intertwiner.
We also require the equivalent operator $R^*(z)$ acting on
$V^*_{z_1}\otimes V_{z_2}$.
It is given as
$R^*(z)=(C\otimes\id)^{-1}R(zq^{-2})(C\otimes\id)$.
Explicitly,
$$
\eqalign{
R^*(z)&v^*_\pm\otimes v_\mp=v^*_\pm\otimes v_\mp,\cr
R^*(z)&v^*_{+}\otimes v_{+}=
  {q-q^{-1}z\over1-z}\,v^*_{+}\otimes v_{+}
 +{q-q^{-1}\over1-z}\,v^*_{-}\otimes v_{-},\cr
R^*(z)&v^*_{-}\otimes v_{-}=
  {q-q^{-1}\over1-z}z\,v^*_{+}\otimes v_{+}
 +{q-q^{-1}z\over1-z}\,v^*_{-}\otimes v_{-}.\cr}
$$

\subsection{Notation}
Our notations for $\slth$ are as follows.
The Cartan subalgebra $\gH$ is spanned by $\{h_0,h_1,d\}$ and
$\alpha_0$, $\alpha_1$ are the roots.
They are related to the fundamental weights by
$\alpha_0=2\Lambda_0-2\Lambda_1+\delta$,
$\alpha_1=2\Lambda_1-2\Lambda_0$;
we also write $\rho=\Lambda_0+\Lambda_1$.
The invariant form on $\gH^*$ is given by
$2(\Lambda_i,\Lambda_j)=\delta_{1i}\delta_{1j}$, $(\Lambda_i,\delta)=1$,
$(\delta,\delta)=0$.
The weight lattice and its dual are
$P=\BZ\Lambda_0\oplus\BZ\Lambda_1\oplus\BZ\delta$ and
$P^*=\BZ h_0\oplus\BZ h_1\oplus\BZ d$, with
$\langle\Lambda_i,h_j\rangle=\delta_{ij}$,
$\langle\Lambda_i,d\rangle=0$, $\langle\delta,h_i\rangle=0$,
$\langle\delta,d\rangle=1$.
We identify $P^*$ with a subset of $P$ via $(\ ,\ )$,
so that  $\alpha_i=h_i$ and
$$
2\rho=4d+h_1.
\eqno(rho)
$$
This plays an important r\^ole in relating $\KX$ and $\KC$ with
$\rho$ and $d$.

There are two irreducible infinite dimensional level $1$ highest
weight $\Uq$ modules,
$V(\Lambda_0)$, $V(\Lambda_1)$, corresponding to the fundamental weights
$\Lambda_0$, $\Lambda_1$.
Their relation to the eigenvectors of the \CTM\ will be discussed in
the next section.

\section{The \BCTM\ and its eigenvectors}

In  \cite{FM92} Foda and Miwa identified a
six-vertex \CTM\ generator with
the derivation $d$ of $\Uq$, and its eigenvectors with the weight
vectors of the level-$1$ modules.
There are two bare ground states,
$(\cdots,-,+,-,+)$ and $(\cdots,+,-,+,-)$.
They defined the action
of the algebra on the sets of paths ${\cal P}_i$, $(i=0,1)$, which
differ from these states at only a finite  number of places.
The problem is to show that everything works consistently
in each order of perturbation.
Convincing evidence was given for this, through extensive
analytic computations to quite high powers of $q$, but the main
result remained a conjecture.
Moreover the $q$-expansions about these bare states are
divergent, and an infinite renormalisation is involved.

In the next two subsections we explain briefly the approach of
Davies \cite{Dav93},
which uses finite size truncation together with the Kashiwara's crystal
base theory \cite{Ka90,Ka91} to prove the  necesary identifications.
This method generalises quite easily to the higher level and higher rank
cases.

\subsection{Identification of derivation}
$\KX$ is defined using
the Boltzmann weights $W$, but it is convenient to introduce another
generator $\KC$ defined from
$\overline{W}$.  It is built from two-site operators $H_j$:
$$
H_j= {1\over2(q-q^{-1})}\left(
 \sigma_{j+1}^x\sigma_j^x+\sigma_{j+1}^y\sigma_j^y
 +\Delta\sigma_{j+1}^z\sigma_j^z
  +{(q-q^{-1})\over2}(\sigma_{j+1}^z - \sigma_j^z)\right).
$$
obtained by expansion of $\check{R}$ about $z=1$.  The subscript $j$
indicates that $H_j$ operates at the positions $j+1$,
$j$ of the chain.
For this reason, we attach a second subscript to the generators
$e_i$, $f_i$, $t_i$, writing $e_{i,j}$, $f_{i,j}$, $t_{i,j}$.
The truncated \CTM\ generator is
$$
\KC^{(N)}=\sum_{j=1}^{N-1}j H_j.
$$
It acts on the $N$-fold tensor product $V^{\otimes N}$.
Write $\hat{e}_i$, $\hat{f}_i$, $\hat{t}_i$ for the $N$-fold
iterated co-product of the generators,
$$
\eqalign{
\hat{e}_i&=\sum_{j=1}^N t_{i,N}\otimes\cdots\otimes  t_{i,j+1}\otimes
e_{i,j}\otimes \id_{j-1}\otimes\cdots\otimes\id_1, \cr
\hat{f}_i&=\sum_{j=1}^N \id_N\otimes\cdots\otimes\id_{j+1}\otimes
f_{i,j}\otimes t_{i,j-1}^{-1}\otimes\cdots\otimes t_{i,1}^{-1}, \cr
\hat{t}_i&=\phantom{\sum_{j=1}^N}  t_{i,N}\otimes\cdots\otimes
t_{i,1}. \cr}
$$

We need their commutation relations with $\KC^{(N)}$.
For $\hat{e}_1$, $\hat{f}_1$ this is
trivial, $\check{R}(x,y)$ intertwines the action of the subalgebra $\U$
which does not depend on the spectral parameter.
If follows immediately that $\KC^{(N)}$ is $\U$ invariant:
$$
[\KC^{(N)},\hat{e}_1]=0,\qquad [\KC^{(N)},\hat{f}_1]=0.
$$
The intertwining condition for $e_0$, $f_0$ gives the
commutation relations \cite{Dav93}
$$
\eqalign{[\KC^{(N)},\hat{e}_0]&=\hat{e}_0 -N
e_{0,N}\otimes\id_{N-1}\cdots\otimes\id_1,
\cr [\KC^{(N)},\hat{f}_0]&=-\hat{f}_0 +N f_{0,N}\otimes
t_{0,N-1}^{-1}\otimes\cdots\otimes t_{0,1}^{-1}. \cr}
$$
This is the vital result.
In the infinite $N$ limit the boundary terms
may be neglected, so that
$\KC=\lim_{N\rightarrow\infty}(\KC^{(N)}-R_N)$ acts as the
derivation in the algebra $\Uq$, verifying the relations \ref{CD}.
$R_N$ is a renormalisation constant and plays no further r\^ole.

\subsection{Identification of modules}
$\KX$ and $\KC$ act formally on a semi-infinite tensor product of
spin-half modules, as does the quantum affine algebra.
It must be emphasised that this limit is only formally defined.
When the eigenstates are developed either as
$\ell_2$-sequences or as $q$-expansions,
one obtains divergent expressions.
The situation is similar to the Ising model \CTM s \cite{Dav88,DP90}.
Having truncated the chain at a finite length, the action of the
algebra is well-defined.
$\KC^{(N)}$ is renormalised by its ground state eigenvalue.
It is shown in \cite{Dav93} that the eigenvectors with eigenvalues up to
$M$ carry the $\Uq$ action to within error terms of order
$q^{2N-2M-1}$.
So the boundary terms carry factors of the order $Nq^{\approx 2N}$,
which is why they may be neglected in the infinite limit.
It is also shown there, using the theory of the crystal base, that the
eigenvectors are in one-to-one correspondence with the weight
vectors of the modules $V(\Lambda_i)$.

This method of finite size approximation had already been noted
for the crystal base \cite{JMMO91}, so it is natural to extend it to
$q\neq0$.
The meaning of the approximation is quite precise, and illustrates
the non-uniform nature of the convergence.
For any fixed integer $M$, the eigenstates corresponding to the
grading levels $\le M$ approximate weight vectors up to that grading
level to within errors of order $Nq^{2N-4M-1}$ as $N\rightarrow\infty$.
But if one selects some order of approximation, and then increases $M$
and $N$ together, the approximated eigenvectors become an increasingly
small fraction of the totality of eigenvectors of the truncated \CTM,
since $2^N/2^M$ increases exponentially fast.
What matters is that $q$-expansions of physical
quantities obtained from the \CTM\ and from representation theory will
agree to every order in $q$.

Without further ado, we shall identify the eigenvectors of the \CTM\ with the
weight vectors of the highest weight modules, and the generator $\KC$ with
the derivation operator $d$.
Examination of the gauge term in $\overline{W}$ shows that
$\KX = 2\KC + S$, $2S=\sum_{k=1}^\infty\sigma_j^z$.
Using the connections \ref{rho} leads to the identification
$$
\KX=2d+h_1/2=\rho.
$$

\subsection{Local properties}
Let $L$ be a local operator, acting at the sites $1,\cdots,n$.
The expectation value $\langle L\rangle$ is calculated by the
standard \CTM\ method \cite{Baxbk} as a quotient of traces:
$$
\langle L\rangle
={\tr(LABCD)\over\tr(ABCD)}
={\tr_{V(\Lambda_i)}(Lq^{2\rho})\over\tr_{V(\Lambda_i)}(q^{2\rho})}.
\eqno(trq)
$$
Here $A$, $B$, $C$ and $D$ are \CTM s for the four quadrants of the
system.
The difficulty is that $L$ must be expressed in a basis of
eigenvectors of $\KX$.
Without an explicit construction for these, the calculation is
restricted to an operator which commutes with $\KX$, i.e., a
one-point function.
For the six-vertex model this $L=(-1)^S$ and the one-point function
is the magnetisation in the ``interaction round a face''
representation \cite{Baxbk}.
The traces are specialisations of the character formulae:
$$
\chi_i(x,y)=
\tr_{V(\Lambda_i)}(x^d y^{h_1})=y^{-i}
{(-x^{1-i}y^2;x^2)_\infty(-x^{1+i}y^{-2};x^2)_\infty
\over(x;x^2)_\infty}.
$$
Substituting $x=q^4$, $y=q$ into these formulae gives the
magnetisation as
$$
m ={(q^2;q^4)_\infty\over(-q^2;q^4)_\infty}
=\prod_{n=0}^\infty{(1-q^{4n+2})\over(1+q^{4n+2})}.
$$

\section{Local description of \BCTM\ eigenvectors}

\subsection{Change of basis}
The difficulty of defining the \CTM\ eigenvectors as
normalisable $\ell_2$ sequences has already been noted.
Representation theory provides a way around the problem.
Consider the tensor product $V(\Lambda_{1-i})\otimes V$.
Formally, we  identify this with
$(\otimes_{n=2}^\infty V_n)\otimes V_1$, with boundary condition
for sector $i$, but counted from the first site.
(There is some abuse of notation: $V_n$ here denotes the
spin-half module for site $n$.)
The product may also be written $(\otimes_{n=1}^\infty V_n)$
and then equated to weight vectors of $V(\Lambda_i)$.
The map between the two pictures is a
$q$-vertex operator (type I Vertex Operators or \VO\ \cite{FR92})
of Frenkel and Reshetikhin \cite{FR92}.
$$
\hgrid=10pt\vgrid=10pt\sarrowlength=20pt
\gridcommdiag{
  \cdots&&\hcross&&\hcross&&\hcross&&&&
  \mathop{\longrightarrow}\limits^\sim&&&
  \cdots&&\hcross&&\hcross&&\hcross&&\hcross&&\otimes&\hcross&}
$$
In this diagram crosses represents sites in the chain (copies
of $V$), not vertices in transfer matrices.

\VO s are simply homomorphisms --- they
preserve the  algebra structure.
They may also be interpreted as expansions of the \CTM\ eigenvectors
for boundary sector $i$ in terms of the state variable at site $1$ and the
eigenvectors for the boundary sector $1-i$.
This process may be iterated to obtain information about any finite
number of state variables at sites $1,\ldots,n$.

\VO s give precisely the information required for the
calculation of local properties in \ref{trq}.
Moreover, the expansions have convergent domains in $q$ and $z$.
But the most crucial property is that the intertwining property
of the \VO s uniquely determines them to within normalisation.
Once one has made the identifications of  section 4,
no further physical arguments are required.
The introductory comments above, and those of section $5.3$ below, are
quite superfluous in this sense.
They are included to provide some intuitive understanding of
the physical meaning of the theory.

\subsection{Vertex operators}
The basic definition of \VO s is that they are $\Up$ homomorphisms
$$
\eqalign{
\Phi^{\Lambda_{1-i}V}_{\Lambda_i}&:
V(\Lambda_i)\rightarrow V(\Lambda_{1-i})\otimes V,\cr
\Phi_{\Lambda_{1-i}V}^{\Lambda_i}&:
V(\Lambda_{1-i})\otimes V\rightarrow V(\Lambda_i).\cr}
$$
We discuss $\Phi^{\Lambda_{1-i}V}_{\Lambda_i}$.
It gives an expansion of weight vectors $u\in V(\Lambda_i)$ as tensor
products:
$$
u=\sum_{n\in\BZ\atop j=\pm}w_{n,j}\otimes v_j,\quad
w_{n,j}\in V(\Lambda_{1-i}).
$$
The intertwining property
$$
\Phi^{\Lambda_{1-i}V}_{\Lambda_i}(xu)=
\Delta(x)\Phi^{\Lambda_{1-i}V}_{\Lambda_i} u,
\quad\forall\,x\in\Up,
$$
implies that $\wt(u)=\wt(w_{n,j})+\wt(v_j)+n\delta$.
Plainly speaking, the expansion is over states of all energy whose
spins match.
The normalisation is fixed by setting the coefficient of the term
in the highest weight vectors $u_{\Lambda_i}$ to unity:
$$
\eqalign{
\Phi^{\Lambda_{1-i}V}_{\Lambda_i}&u_{\Lambda_i}
=(u_{\Lambda_{1-i}}\otimes v_{1-i}) + \cdots,\cr
\Phi_{\Lambda_{1-i}V}^{\Lambda_i}&(u_{\Lambda_{1-i}}\otimes v_{1-i})
=u_{\Lambda_i} + \cdots. \cr}
$$

\VO s may also be defined to preserve the grading operator $d$:
they are the $\Uq$ homomorphisms
$$
\eqalign{
\Phit^{\Lambda_{1-i}V}_{\Lambda_i}(z)&:  V(\Lambda_i)\rightarrow
V(\Lambda_{1-i})\otimes V_z,\cr
\Phit_{\Lambda_{1-i}V}^{\Lambda_i}(z)&: V(\Lambda_{1-i})\otimes
V_z\rightarrow V(\Lambda_i).\cr}
$$
The two forms ($\Phi$ and $\Phit(z)$) are trivially related by
inserting or deleting the appropriate powers of $z$,
$$
u=\sum_{n\in\BZ\atop j=\pm}w_{n,j}\otimes z^n v_j.
$$
Since the derivation detects the grading before and after the
action of the \VO, the powers of $z$ may be inserted by conjugation
with the operator $z^d$:
$$
\Phit^{\Lambda_{1-i}V}_{\Lambda_i}(z)=
z^{d\otimes\id}\,\Phi^{\Lambda_{1-i}V}_{\Lambda_i}z^{-d}.
$$
{}From this follows  the  ``homogeneity property''
$$
\eqalign{
\Phit_{\Lambda_i}^{\Lambda_{1-i}V}(z_1z_2)=
z_1^{d\otimes\id}\,\Phit_{\Lambda_i}^{\Lambda_{1-i}V}(z_2)z_1^{-d}, \cr
\Phit^{\Lambda_i}_{\Lambda_{1-i}V}(z_1z_2)=
z_1^d\Phit^{\Lambda_i}_{\Lambda_{1-i}V}(z_2)z_1^{-d\otimes\id}. \cr}
$$

Some elementary calculations give the following results:
$$
\eqalignno{
&\Phit_{\Lambda_0}^{\Lambda_1 V}(z) u_{\Lambda_0}=
u_{\Lambda_1}\otimes v_{-}
-q(f_1 u_{\Lambda_1})\otimes v_{+}+{\cal O}(z), \cr
&\Phit_{\Lambda_1}^{\Lambda_0 V}(z) u_{\Lambda_1}=
u_{\Lambda_0}\otimes v_{+}+{\cal O}(z), &(int)\cr
&\Phit_{\Lambda_1}^{\Lambda_0 V}(z) (f_1u_{\Lambda_1})=
u_{\Lambda_0}\otimes v_{-}+{\cal O}(z). \cr}
$$
They will be used below to fix normalisations.

\subsection{Physical description}
Let $P^{\gamma,j}_\beta$ be a non normal\-ised
semi-infinite shift operator with entries
$$
P_{\ldots,\beta_3,\beta_2,\beta_1}^{\ldots,\gamma_2,\gamma_1,j}
\simeq\cdots\delta_{\beta_3,\gamma_2}
\delta_{\beta_2,\gamma_1}\delta_{\beta_1,j}.
$$
Since $P$ preserves the $\Up$ action
one may identify an appropriate constant multiple with the \VO\
$\Phi_{\Lambda_i}^{\Lambda_{1-i}V}$.
Further, we use the relation of $d$ with the \CTM\ to write
$\Phit_{\Lambda_i}^{\Lambda_{1-i}V}(z)=A_{1-i}(z)PA_i(z^{-1})$.
We may even identify the \VO\ as a normalised row transfer matrix.
This is a generalisation of the boost property
for the full transfer matrix \cite{SW83,Th86}.
The product $PA_i$ is the partition function of a
quadrant with the boundary type $i$ along its horizontal edge and $1-i$
along its vertical one.
It is not a \CTM, but may be re-written as a different product:
$$
\hgrid=10pt\vgrid=10pt\sarrowlength=20pt
\hbox{$\gridcommdiag{
  &&\hskip5ptA_i&&\hcross&&&\blab3&\hcross&&&\clab3\cr\cr
  &&\hcross&&\hcross&&&\blab2&\hcross&&&\clab2\cr\cr
  \hcross&&\hcross&&\hcross&&&\blab1&\hcross&&&\clab1\cr\cr
  &\alab3&&\alab2&&\alab1&&\hskip5pt\raise5pt\hbox{$P$}&\hskip10pt
j\cr}$} ~=~
\hbox{$\gridcommdiag{
  &&A_{1-i}&&\hcross&&&\clab2\cr\cr
  &&\hcross&&\hcross&&&\clab1\cr\cr
  &\blab3&&\blab2&&\blab1\cr\cr
  \hcross&&\hcross&&\hcross&&\hskip10pt j\cr\cr
  \hskip-10pt\raise5pt\hbox{$\Phit$}&\alab3&&\alab2&&\alab1\cr\cr }$}
$$
The semi-infinite row of vertices corresponds to the vertex operator
$\Phit^{\Lambda_{1-i}V}_{\Lambda_i}(z)$, with a normalisation
factor $\phi_i(z)$.
$$
\def\cross{\hskip-7.1pt\lower2.5pt\hbox{$\hcross$}}
\hgrid=10pt\vgrid=10pt\sarrowlength=20pt
\gridcommdiag{
\cross&&\cross&&\cross&&&&&&&
\hskip3pt\lower2pt\hbox{$=\quad\phi_i(z)
\Phit_{\Lambda_i}^{\Lambda_{1-i}V}(z).$}&&&&\cr}
$$
This physical interpretation of  \VO s follows \cite{FJMMN93}.

The product of row transfer matrices is
commuted using the Yang Baxter equations together
with  the assumption that boundary effects disappear for a
system with fixed boundary conditions:
$$
\def\cross{\hskip-7.1pt\lower2.5pt\hbox{$\hcross$}}
\hgrid=10pt\vgrid=10pt\sarrowlength=20pt
\gridcommdiag{
\cross&&\cross&&\cross&&&&\lower3pt\hbox{$z_2$}
&&&&&&\cross&&\cross&&\cross&&\hskip5pt\lower3pt\hbox{$z_2$}\cr
&&&&&&\dcross&&&&&
\lower2.5pt\hbox{$=$}&&&&&&&\cr
\cross&&\cross&&\cross&&&&\lower3pt\hbox{$z_1$}
&&&&&&\cross&&\cross&&\cross&&\hskip5pt\lower3pt\hbox{$z_1$}\cr}
$$
We use this to fix normalisation factors.
Write
$$
\check{R}(z_1/z_2)
 \Phit_{\Lambda_{1-i}}^{\Lambda_iV}(z_1)
 \Phit_{\Lambda_i}^{\Lambda_{1-i}V}(z_2)
=
 \kappa(z_1/z_2)
 {\phi_{1-i}(z_2)\phi_i(z_1)\over
  \phi_{1-i}(z_1)\phi_i(z_2)}
 \Phit_{\Lambda_{1-i}}^{\Lambda_iV}(z_2)
 \Phit_{\Lambda_i}^{\Lambda_{1-i}V}(z_1),
$$
and take the vacuum expectation.
{}From the definitions, the terms in
$\Phit_{\Lambda_i}^{\Lambda_{1-i}V}(z_1)u_{\Lambda_i}$
involve only non-negative powers of $z_1$.
In order to arrive
once more at the  highest weight vector after the action of
$\Phit_{\Lambda_{1-i}}^{\Lambda_iV}(z_2)$, these terms must pair with a
matching negative power of $z_2$.
Denote the vacuum expectation as
$$
\Psi_i(z_1/z_2)=
\langle u_{\Lambda_i}\mid
\Phit_{\Lambda_{1-i}}^{\Lambda_iV}(z_1)
\Phit_{\Lambda_i}^{\Lambda_{1-i}V}(z_2)
\mid u_{\Lambda_i}\rangle.
$$
It involves only non-negative powers of $z_1/z_2$, and
will be analytic in $z_1/z_2$ in some domain containing the origin.
As an element of $V\otimes V$ it must be an  eigenvector of
$\check{R}(z_1/z_2)$ because of the commutation relation.
A simple calculation gives:
$$
\eqalign{
\check{R}(z)&(v_{+}\otimes v_{-} - q v_{-}\otimes v_{+})
={z-q^2\over1-q^2 z}\,
(v_{+}\otimes v_{-} - q v_{-}\otimes v_{+}), \cr
\check{R}(z)&(v_{-}\otimes v_{+} - q z^{-1} v_{+}\otimes v_{-})
={1-q^2 z^{-1}\over1-q^2 z}\,
(v_{-}\otimes v_{+} - qz^{-1} v_{+}\otimes v_{-}). \cr}
$$
To leading order this corresponds exactly with \ref{int}.
Inserting the formula for $\kappa(z)$ and using analyticity,
$$
\eqalignno{
\Psi_0(z)&=
{(q^6z;q^4)_\infty\over(q^4z;q^4)_\infty}\,
(v_{+}\otimes v_{-} - q v_{-}\otimes v_{+}), &(norm)\cr
\Psi_1(z)&=
{(q^6z;q^4)_\infty\over(q^4z;q^4)_\infty}\,
(v_{-}\otimes v_{+} - qz^{-1} v_{+}\otimes v_{-}). \cr}
$$
and
$$
\phi_i(z)=\gamma_i z^{\Delta_{1-i}-\Delta_i},\qquad
\Delta_0=0,\qquad
\Delta_1=1/4,
$$
for some constants $\gamma_i$.
The ``conformal factors'' $\Delta_i$, which were included in the
definition of the \VO s in \cite{DFJMN92}, are a shift in the
zero point; $A_i(z)\simeq z^{d+\Delta_i}$.
Such shifts are calculated for the Ising model in \cite{DP90}.

\subsection{Mathematical construction}
We have discussed the physical construction of the \VO s in some detail.
However, for the purpose of calculation, the mathematical construction
is preferable.
The scale factors \ref{norm} are determined unambiguously in
\cite{DFJMN92,DO93} by solving the $q$-KZ (Knihnik-Zamolodchikov)
equation \cite{FR92}.
{}From this, the commutation relation for the \VO s follows
immediately, viz:
$$
\eqalign{
PR(z_1/z_2)
 \Phit_{\Lambda_{1-i}}^{\Lambda_iV}(z_1)
 \Phit_{\Lambda_i}^{\Lambda_{1-i}V}(z_2)
 &=z^{\delta_{i0}} r(z_1/z_2)
 \Phit_{\Lambda_{1-i}}^{\Lambda_iV}(z_2)
 \Phit_{\Lambda_i}^{\Lambda_{1-i}V}(z_1), \cr
r(z)&=
 {(zq^4;q^4)_\infty(z^{-1}q^6;q^4)_\infty\over
  (z^{-1}q^4;q^4)_\infty(zq^6;q^4)_\infty}. \cr}
$$
This confirms our identification of the conformal factors, and
the r\^ole of the partition function, in interpreting the
\VO s as transfer matrices.

The charge conjugation isomorphism $C$ relates
$\Phit_{\Lambda_i}^{\Lambda_{1-i}V^*}(z)$ to
$\Phit_{\Lambda_i}^{\Lambda_{1-i}V}(z)$ by
$$
\Phit_{\Lambda_i}^{\Lambda_{1-i}V^*}(z)=
(-q)^{\delta_{i1}}(\id\otimes C)^{-1}
\Phit_{\Lambda_i}^{\Lambda_{1-i}V}(zq^{-2}),
$$
and from this the commutation relation emerges in a form
needed for the calculation of the spontaneous polarisation.
It is
$$
PR^*(z)
 \Phit_{\Lambda_{1-i}}^{\Lambda_iV^*}(z_1)
 \Phit_{\Lambda_i}^{\Lambda_{1-i}V}(z_2)
 =-q^{-1}z^{\delta_{i0}} r(zq^{-2})
 \Phit_{\Lambda_{1-i}}^{\Lambda_iV}(z_2)
 \Phit_{\Lambda_i}^{\Lambda_{1-i}V^*}(z_1).
$$

To complete the determination of local operators, we must construct
the inverse of $\Phit_{\Lambda_i}^{\Lambda_{1-i}V}(z)$.
By definition, it is
$\Phit^{\Lambda_i}_{\Lambda_{1-i}V}(z)$
up to normalisation.
The construction goes as follows.
The \VO s are algebra homomorphisms, so one may use the canonical
identification that
$\Hom(L\otimes N,M)$ is isomorphic to $\Hom(L,M\otimes N^*)$,
to equate $\Phit^{\Lambda_i}_{\Lambda_{1-i}V}(z)$ with
$\Phit_{\Lambda_i}^{\Lambda_{1-i}V^*}(z)$.
Let us expand on this.
It means that
$$
\Phit_{\Lambda_i V}^{\Lambda_{1-i}}(z)(u\otimes v_j)=
\langle\Phit_{\Lambda_i}^{\Lambda_{1-i}V^*}(z)u,
v_j\rangle,
$$
where the pairing is in $V^*\otimes V$.
One may check that this is correct on the Chevalley generators.
For example,
$$
\eqalign{
\Phit_{\Lambda_i V}^{\Lambda_{1-i}}(z)&(\Delta(e_i)(u\otimes v_j))=
\langle e_i\Phit_{\Lambda_i}^{\Lambda_{1-i}V^*}(z)u,v_j\rangle \cr
&+\sum_k\left(t_i\Phit_{\Lambda_i}^{\Lambda_{1-i}V^*}(z)u\right)_k
\Big(\langle e_i v^*_k,v_j\rangle
+\langle t_i v^*_k,e_i v_j\rangle\Big). \cr}
$$
The second line vanishes because the antipode is used to define
the dual.
Finally, \ref{norm} tells us the scalar factor for the inverse.
Setting $z=q^{-2}$ gives
$$
\Phi^{\Lambda_i}_{\Lambda_{1-i}V}(z)
\Phi_{\Lambda_i}^{\Lambda_{1-i}V}(z)
={(q^4;q^4)_\infty\over(q^2;q^4)_\infty}\,\id_{V(\Lambda_i)}.
$$
This completes the construction of change of basis matrices.

\subsection{Spontaneous polarisation}
\quad Two decades ago, Baxter calculated the spontaneous polarisation of the
six-vertex model using the Bethe Ansatz \cite{Bax73}.
Unlike many other important results, no alternative derivation was
subsequently forthcoming, although the method has recently been extended to
the calculation of the mean polarisation in the higher spin case
\cite{DJMO91}.
The properties of \VO s gives the first new derivation: moreover \VO s give
a general method to obtain correlation functions.
The ideas have even been used for the eight-vertex model \cite{JMN92},
validating a long-standing conjecture \cite{BK74}, although the underlying
algebraic structure is not presently understood in that case.

Here we sketch the derivation for the six-vertex model, which is rank-$1$.
The solution for arbitrary rank has recently been given by Koyama
\cite{K93}.
The expectation value that a single edge variable has the value
$m$ ($=\pm$) is simply the trace of an operator with matrix elements
$E_{jk}=\delta_{jm}\delta_{km}$.
Recalling  our identification
of the \CTM\ generator $\KC$ with $\rho$,
consider the more general unnormalised expectation value
$$
F^{(i)}(z_1/z_2)=
\tr_{V(\Lambda_i)}(q^{-2 \rho}
 \Phit^{\Lambda_i V^*}_{\Lambda_{i-1}}(z_1)
 \Phit^{\Lambda_{1-i} V}_{\Lambda_i}(z_2)).
$$
It takes its values in $V^*\otimes V$
and the spontaneous polarisation is the ratio
$(F^{(i)}_{++}(1)-F^{(i)}_{--}(1))/(F^{(i)}_{++}(1)+F^{(i)}_{--}(1))$.

We may derive a $q$-difference equation for $F^{(i)}$:
$$
\eqalign{
F^{(i)}(z_1&/z_2 q^4) \cr
= &\tr_{V(\Lambda_i)}(q^{-2\rho}
 \Phit^{\Lambda_i V^*}_{\Lambda_{1-i}}(z_1)
 \Phit^{\Lambda_{1-i} V}_{\Lambda_i}(z_2 q^4)) \cr
= &P\,\tr_{V(\Lambda_{1-i})}
 (\Phit^{\Lambda_{1-i} V}_{\Lambda_i}(z_2 q^4)
 q^{-2\rho}\Phit^{\Lambda_i V^*}_{\Lambda_{1-i}}(z_1)) \cr
= &P\,(q^{-h_1}\otimes\id)
 \times\tr_{V(\Lambda_{1-i})}
 (q^{-2\rho}\Phit^{\Lambda_{1-i} V}_{\Lambda_i}(z_2)
 \Phit^{\Lambda_i V^*}_{\Lambda_{1-i}}(z_1)) \cr
= &-qz^{-\delta_{i1}} r(zq^{-2})
 (\id\otimes q^{-h_1})R^*(z)
 \times\tr_{V(\Lambda_{1-i})}(q^{-2\rho}
 \Phit^{\Lambda_{1-i} V^*}_{\Lambda_i}(z_1)
 \Phit^{\Lambda_i V}_{\Lambda_{1-i}}(z_2)). \cr}
$$
The three steps use, respectively, the cyclic property of the trace,
the homogeneity and intertwining property of the \VO s, and the
commutation relations.
The equation thus obtained reads
$$
F^{(i)}(z q^{-4}) = -qz^{-\delta_{i1}}
 (\id\otimes q^{-h_1})r(zq^{-2}) R^*(z) F^{(1-i)}(z).
$$

These equations may be solved quite simply.
Set $G^{(\pm)}(\zeta)=F^{(1)}(\zeta^2)\pm q\zeta F^{(0)}(\zeta^2)$,
which decouples them.
Note also that only the 2 by 2 block  spanned
by $v^*_\pm\otimes v_\pm$ is non zero.
For it we find
$$
\eqalign{
&G^{(\pm)}(\zeta q^{-2})=
\mp r(\zeta^2 q^{-2}) M(\zeta) G^{(\pm)}(\zeta), \cr
&M(\zeta)={\zeta\over1-\zeta^2}
\pmatrix{1-q^{-2}\zeta^2&\zeta^2(1-q^{-2})\cr
         q^2-1&q^2-\zeta^2}. \cr}
$$
This may be reduced to scalar equations by considering a
generalised eigenvalue problem:
$M(\zeta) w(\zeta)=\lambda(\zeta)w(\zeta q^{-2})$.
The eigenvectors are
$$
{w_\pm(\zeta)\over(1\mp\zeta)}=(1,\pm\zeta^{-1}),\quad
\lambda(\zeta)=\zeta.
$$
The solution for $G^{(\pm)}(\zeta)$ involves $\Theta_{q^2}(\pm\zeta)$, where
$$
\Theta_p(\zeta)=(p;p)_\infty (\zeta;p)_\infty (\zeta^{-1} p;p)_\infty.
$$
It satisfies $\Theta_p(\zeta p)=-\zeta^{-1}\Theta_p(\zeta)$.
In order to have the required analyticity a zero of this
function must cancel the factor $(1\mp\zeta)$ in $w_\pm(\zeta)$
and this implies that there is only one non-zero component
for each of $G^{(\pm)}(\zeta)$.
The argument whereby one shows that $F^{(i)}(z)$ has the assumed
analyticity is based on finding an integral representation,
and is given in \cite{JMN92,K93}.
The solutions may be normalised by the fact that
$F^{(i)}(q^{-2})$ is the trace of the identity operator.
However, this is not necessary for finding the staggered polarisation.
The $G^{(\pm)}(\zeta)$ differ only in the factors
$\Theta_{q^2}(\pm\zeta)$, and we find, independent of $i$,
$$
P
 =\left[{(1+\zeta)\Theta_{q^2}(\zeta)\over
   (1-\zeta)\Theta_{q^2}(-\zeta)}\right]_{\zeta=1}
 =\prod_{n=0}^\infty
 {(1-q^{2n+2})^2\over(1+q^{2n+2})^2}.
$$

The beautiful interplay which has
emerged between physics and mathematics
brings to mind a quote from the writings of Ludwig Boltzmann:

{\narrower\noindent\it It is unbelievable how simple and straight forward
each result appears once it has been found, and how difficult it seems   so
long as the way which leads to it is unknown.
\par}

\section*{Acknowledgements}

I wish to acknowledge many helpful discussions with
D. Altschuler,
M. T. Batchelor,
R. J. Baxter,
A. Carey,
O. Foda,
M. Jimbo,
T. Miwa,
M. Okado,
P. A. Pearce,
I. Peschel,
N. Yu. Reshetikhin,
R. W. Richardson
and R. Zhang.

\section*{Appendix: Some numerical results}

For making finite size numerical calculations it is convenient to use
$\KC^{(N)}$ rather than $\KX^{(N)}$
because the gauge term forces the boundary conditions to
those appropriate for $V(\Lambda_0)$ when $N$ is even and
$V(\Lambda_1)$ when $N$ is odd \cite{Dav93}.
Here we show some results for chains of length up to $11$.
Energies are relative to the antiferromagnetic ground state energy.
They should be compared with the figures for the
weight space structure of the relevant modules.
$\U$-invariance means that the $\hat{f}_1$ multiplets are exact.
The $\hat{f}_0$ multiplets tend to the proper degeneracy
as $N$ increases.
Numerical checks have also been made on the actions of $\hat{e}_0$ and
$\hat{f}_0$ for these values of $N$.
In the weight space diagrams for the level-$1$ modules,
spin is in the horizontal direction and grading is downwards.
The action of $f_1$ is shown as a left pointing arrow, since $f_1$ decreases
the weight by $\alpha_1$.
$f_0$ is a diagonal arrow down and to the right, since $f_0$ decreases the
weight by $\alpha_0$, and $\alpha_0+\alpha_1=\delta$.
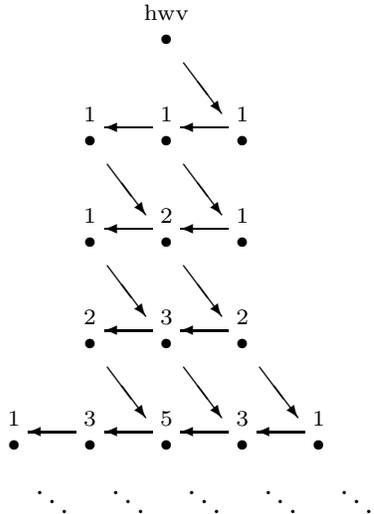
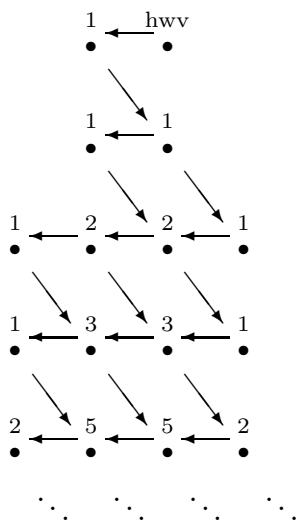
\smallskip
$$
\eqalign{
\def\ws#1{#1\atop\bullet}
\def\darr{\hskip-13pt\lower1pt\hbox{$\drawvector(3,-4){12pt}$}}
\def\harr{\raise2.5pt\hbox{$\drawvector(-1,0){15pt}$}}
\hgrid=6pt\vgrid=8pt
\gridcommdiag{&&&&&&&&\ws{{\rm hwv}}\cr
&&&&&&&&&&&\darr\cr\cr\cr
&&&&\ws1&&\harr&&\ws1&&\harr&&\ws1\cr
&&&&&&&\darr&&&&\darr\cr\cr\cr
&&&&\ws1&&\harr&&\ws2&&\harr&&\ws1\cr
&&&&&&&\darr&&&&\darr\cr\cr\cr
&&&&\ws2&&\harr&&\ws3&&\harr&&\ws2\cr
&&&&&&&\darr&&&&\darr&&&&\darr\cr\cr\cr
\ws1&&\harr&&\ws3&&\harr&&\ws5&&\harr&&\ws3&&\harr&&\ws1\cr\cr\cr
&&\ddots&&&&\ddots&&&&\ddots&&&&\ddots&&&&\ddots}}\quad
\eqalign{
\font\srm cmr10 scaled 900
\def\dash{\hbox{\rm---}}
\def\vline#1{ height #1pt &\omit &&\omit &&\omit &&\omit &&\omit &\cr}
\def\hline{\noalign{\hrule}\vline1}
{\vbox{\offinterlineskip
\hrule
\halign{&\vrule#&\strut~{\srm#}~&\vrule#&~{\srm#}~&\vrule#&~{\srm#}~&\vrule#
&~{\srm#}~&\vrule#&~{\srm#}~&\vrule#\cr
\vline1
           &{\rm Spin}&&$N=4$&&$N=6$     &&$N=8$     &&$N=10$    &\cr
\hline
           &0    &&0         &&0         &&0         &&0         &\cr
\hline
           &1    &&1.0021201 &&1.0002473 &&1.0000301 &&1.0000038 &\cr
\hline
           &0    &&2.1189021 &&2.0203449 &&2.0033092 &&2.0005036 &\cr
           &1    &&2.1167936 &&2.0200996 &&2.0032794 &&2.0004998 &\cr
\hline
           &0    &&\dash     &&3.2770532 &&3.0678881 &&3.0151265 &\cr
           &1    &&3.1200455 &&3.0205449 &&3.0033418 &&3.0005085 &\cr
           &1    &&\dash     &&3.2768064 &&3.0678580 &&3.0151227 &\cr
\hline
           &0    &&\dash     &&4.2968794 &&4.0711308 &&4.0156211 &\cr
           &0    &&\dash     &&\dash     &&4.4835212 &&4.1520062 &\cr
           &1    &&\dash     &&4.2771017 &&4.2771017 &&4.0151267 &\cr
           &1    &&\dash     &&\dash     &&4.4839116 &&4.1520024 &\cr
           &2    &&\dash     &&4.0205469 &&4.0033421 &&4.0005086 &\cr}
\hrule}}}
$$
\centerline{{\bf Weight spaces of $V(\Lambda_0)$ and
negative eigenvalues of truncated $\BCTM$}}
\medskip
$$
\eqalign{
\def\ws#1{#1\atop\bullet}
\def\darr{\hskip-13pt\lower1pt\hbox{$\drawvector(3,-4){12pt}$}}
\def\harr{\raise2.5pt\hbox{$\drawvector(-1,0){15pt}$}}
\hgrid=6pt\vgrid=8pt
\gridcommdiag{
&&&&&&\ws1&&\harr&&\ws{{\rm hwv}}\cr
&&&&&&&&&\darr\cr\cr\cr
&&&&&&\ws1&&\harr&&\ws1\cr
&&&&&&&&&\darr&&&&\darr\cr\cr\cr
&&\ws1&&\harr&&\ws2&&\harr&&\ws2&&\harr&&\ws1\cr
&&&&&\darr&&&&\darr&&&&\darr\cr\cr\cr
&&\ws1&&\harr&&\ws3&&\harr&&\ws3&&\harr&&\ws1\cr
&&&&&\darr&&&&\darr&&&&\darr\cr\cr\cr
&&\ws2&&\harr&&\ws5&&\harr&&\ws5&&\harr&&\ws2\cr\cr\cr
&&&&\ddots&&&&\ddots&&&&\ddots&&&&\ddots&&\cr\cr   }}\quad
\eqalign{
\font\srm cmr10 scaled 900
\def\dash{\hbox{\rm---}}
\def\vline#1{ height #1pt &\omit &&\omit &&\omit &&\omit &&\omit &\cr}
\def\hline{\noalign{\hrule}\vline1}
{\vbox{\offinterlineskip
\hrule
\halign{&\vrule#&\strut~{\srm#}~&\vrule#&~{\srm#}~&\vrule#&~{\srm#}~&\vrule#
&~{\srm#}~&\vrule#&~{\srm#}~&\vrule#\cr
\vline1
           &{\rm Spin}&&$N=5$&&$N=7$     &&$N=9$     &&$N=11$    &\cr
\hline
           &1/2  &&0         &&0         &&0         &&0         &\cr
\hline
           &1/2  &&1.0081308 &&1.0011185 &&1.0001513 &&1.0000204 &\cr
\hline
           &1/2  &&2.1899754 &&2.0396425 &&2.0076092 &&2.0013141 &\cr
           &3/2  &&2.0082121 &&2.0011297 &&2.0001528 &&2.0000206 &\cr
\hline
           &1/2  &&3.1980888 &&3.0407572 &&3.0077598 &&3.0013343 &\cr
           &1/2  &&\dash     &&3.3752382 &&3.1052883 &&3.0268467 &\cr
           &3/2  &&3.1899870 &&3.0396431 &&3.0076093 &&3.0013141 &\cr
\hline
           &1/2  &&4.2000343 &&4.0411612 &&4.0078363 &&4.0013475 &\cr
           &1/2  &&\dash     &&4.3763556 &&4.1054395 &&4.0268671 &\cr
           &1/2  &&\dash     &&\dash     &&4.6000763 &&4.2077221 &\cr
           &3/2  &&4.1999428 &&4.0411497 &&4.0078347 &&4.0013473 &\cr
           &3/2  &&\dash     &&4.3752396 &&4.1052884 &&4.0268467 &\cr}
\hrule}}}
$$
\centerline{{\bf Weight spaces of $V(\Lambda_1)$ and
negative eigenvalues of truncated $\BCTM$}}

\def\PTRSL{Phil. Trans. Roy. Soc. Lond.}
\def\SSR{Sov. Sci. Rev. Math. Phys.}
\section*{References}

\refis{ABF84} \jnlitem
{G. E. Andrews, R. J. Baxter and P. J. Forrester:
Eight-vertex SOS model and generalized Rogers-Ramanujan type identities}
{\JSP}{35}{(1984), 193-266}

\refis{Baxbk} \bkitem
{R. J. Baxter}{Exactly Solved  Models in Statistical Mechanics}
{Academic Press (1982)}

\refis{Bax71} \jnlitem
{R. J. Baxter: Partition function of the eight-vertex lattice model}
{\APNY}{70}{(1972), 193-228}

\refis{Bax73} \jnlitem
{R. J. Baxter:
Spontaneous staggered polarization of the $F$ model}
{\JSP}{9}{(1973), 145-182}\eject

\refis{Bax76} \jnlitem
{R. J. Baxter: Corner transfer matrices of the eight vertex model.
Low temperature expansions and conjectured properties}
{\JSP}{15}{(1976), 485-503}

\refis{Bax78} \jnlitem
{R. J. Baxter:
Solvable eight-vertex model on an arbitrary planar lattice}
{\PTRSL}{289}{(1978), 315-346}

\refis{Bax80} \jnlitem
{R. J. Baxter:
Hard Hexagons: exact solution}
{\JPA}{13}{(1980), L61-L70}
\jnlitem{--- :
Rogers-Ramanujan identities in the hard hexagon model}
{\JSP}{26}{(1981), 427-452}

\refis{Bax81} \jnlitem
{R. J. Baxter : Corner transfer matrices}
{\PA}{106}{(1981), 18-27}

\refis{BK74} \jnlitem
{R. J. Baxter and S. B. Kelland:
Spontaneous  polarization of the eight vertex model}
{\JPC}{7}{(1974), L403-406}

\refis{BPZ84} \jnlitem
{A. A. Belavin, A. M. Polyakov and A. B. Zamol\-odchikov:
Infinite~conformal~symmetry~in two dimensional quantum field theory}
{\NPB}{241}{(1984), 333-380}

\refis{Car84} \jnlitem
{J. L. Cardy: Conformal invariance and surface critical behaviour}
{\NPB}{240}{\hbox{(1984),} 514-532}

\refis{D87} \bkitem
{V.G. Drinfeld}
{Quantum groups, Proc. ICM Berkeley}{(1986), 798-820}

\refis{Dav88} \jnlitem
{B. Davies: Corner transfer matrices for the Is\-ing model}
{\PA}{154}{(1989), 1-20}

\refis{Dav93} \jnlitem
{B. Davies:
Corner transfer matrices and quantum affine algebras}
{\JPA}{}{(1993), in press}

\refis{DJMO91} \jnlitem
{E. Date, M. Jimbo, K. Miki and M. Okado:
Mean staggered polarization for the higher spin analog of the
6-vertex model}
{\IJMPA}{7}{{\it Suppl.\/} 1A, (1992), 151-160}

\refis{DFJMN92} \jnlitem
{B. Davies, O. Foda, M. Jimbo, T. Miwa and A. Nakayashiki:
Diagonalization of the XXZ Hamiltonian by vertex operators}
{\CMP}{151}{(1993), 89-153}

\refis{DJKMO87} \jnlitem
{E. Date, M. Jimbo, A. Kuniba, T. Miwa and M. Okado:
Exactly solvable SOS models: Local height probabilities and
theta function identities}
{\NPB}{290}{(1987), 231-273}

\refis{DJKMO89} \jnlitem
{E. Date, M. Jimbo, A. Kuniba, T. Miwa and M. Okado:
One dimensional configuration sums in vertex models and
affine Lie algebra characters}
{\LMP}{17}{(1987), 69-77}

\refis{DO93} \jnlitem
{E. Date and M. Okado: Calculation of excitation spectra of the
spin model related with the vector representation of the
quantized aff\-ine algebra of type $A^{(1)}_n$}
{Osaka Univ. Math. Sci. Preprint}{1}{(1993)}

\refis{DP90} \jnlitem{B. Davies and  P. A. Pearce:
Conformal invariance and critical spectrum of corner transfer matrices}
{\JPA}{23}{(1990) 1295-1312}

\refis{Fad81} \jnlitem
{L. D. Faddeev:
Quantum completely integrable models in field theory}
{\SSR}{C1}{(1980), 107-155}

\refis{FC91} \jnlitem
{G. Felder and A. LeClair: Restricted quantum affine symmetry
of restricted minimal conformal models}
{\IJMPA}{7}{{\it Suppl.\/} 1A, (1992), 239-278}

\refis{FJMMN93} \jnlitem
{O. Foda, M. Jimbo, K. Miki, T. Miwa and A. Nakayashiki:
Vertex operators in solvable lattice models}{\JMP}{}{to appear}

\refis{FM92} \jnlitem
{O. Foda and T. Miwa:
Corner transfer matrices and quantum affine algebras}
{\IJMPA}{7}{{\it Suppl.\/} 1A, (1992), 279-302}

\refis{FR92} \jnlitem
{I. B. Frenkel and N. Yu. Reshetikhin:
Quantum affine algebras and holonomic difference equations}
{\CMP}{149}{(1992), 1-60}

\refis{FQS84} \bkitem
{D. Friedan, Z. Qiu and S. Shenker}
{Vertex operators in Mathematical Physics,
eds. J. Lepowski, S. Mandelstam and I. M. Singer}
{Springer (1984)}

\refis{GKO85} \jnlitem
{P. Goddard, A. Kent and D. Olive:
Virasoro algebra and coset space models}
{\PLB}{152}{(1985), 88-92}
\jnlitem
{--- : Unitary representations of the Virasoro and
super-Virasoro algebras}{\CMP}{103}{(1986), 105-119}

\refis{IIJMNT92} \jnlitem
{M. Idzumi, K. Iohara, M. Jimbo, T. Miwa, T. Nakayashima and
T. Tokihiro: Affine symmetry in vertex models}
{\IJMPA}{8}{(1993), 1479-1511}

\refis{J85} \jnlitem
{M. Jimbo:
A $q$-difference analogue of $U(\goth g)$ and the Yang-Baxter equation}
{\LMP}{10}{(1985), 63-69}

\refis{J89} \jnlitem
{M. Jimbo: Introduction~~to~~the~~Yang-Baxter equation}
{\IJMPA}{4}{(1989), 3759-3777}

\refis{J92} \bkitem
{M. Jimbo}
{Nankai Lecture Notes on Mathematical Physics, ed. Ge Mo-Lin}
{World Scientific (1992)}

\refis{JMMO91} \jnlitem
{M. Jimbo, K.C. Misra, T. Miwa and M. Oka\-do:
Combinatorics of representations of $U_{q}(\hat sl(n))$ at $q=0$}
{\CMP}{136}{(1991), 543}

\refis{JMO88a} \jnlitem
{M. Jimbo, T. Miwa and M. Okado:
Solvable lattice models related to the vector representation of
classical simple Lie algebras}
{\CMP}{116}{(1988), 507-525}

\refis{JMO88b} \jnlitem
{M. Jimbo, T. Miwa and M. Okado:
Local state probabilities of solvable lattice models:
An $A^{(1)}_{n-1}$ family}
{\NPB}{300}{(1988), 74-108}

\refis{JMMO91} \jnlitem
{M. Jimbo, K.C. Misra, T. Miwa and M. Okado:
Combinatorics of representations of $U_{q}(\hat sl(n))$ at $q=0$}
{\CMP}{136}{(1991), 543-566}

\refis{JMO92} \jnlitem
{M. Jimbo, T. Miwa and Y. Ohta:
Structure of the space of states in RSOS models}
{\IJMPA}{8}{(1993), 1457-1477}

\refis{JMMN92} \jnlitem
{M. Jimbo, K. Miki, T. Miwa and A. Nakaya\-shiki:
Correlation Functions of the \XXZ\ mod\-el for $\Delta<-1$}
{\PLA}{168}{(1992), 256-263}

\refis{JMN92} \jnlitem
{M. Jimbo, T. Miwa and A.  Nakayashiki:
Difference equations for the correlation functions
of the eight-vertex model}
{\JPA}{26}{(1993), 2199-2209}

\refis{Ka90} \jnlitem
{M. Kashiwara:
Crystalizing the $q$-analogue of universal enveloping algebras}
{\CMP}{133}{(1990), 249-260}

\refis{Ka91} \jnlitem
{M. Kashiwara:
On crystal bases of the $q$-analog\-ue of universal enveloping algebras}
{\Duke}{63}{(1991), 465-516}

\refis{(KMN)^2} \jnlitem
{S.J. Kang, M. Kashiwara, K.C. Misra, T. Miwa, T. Nakashima and
A. Nakayashiki: Affine crystals and vertex models}
{\IJMPA}{7}{{\it Suppl.\/} 1A, (1992), 449-484}

\refis{K93} \jnlitem
{Y. Koyama:
Staggered Polarization of Vertex models with $U_p(\gs\gl(n))$ Symmetry}
{\CMP}{}{(1993), in press}

\refis{LL63} \jnlitem
{E. H. Lieb and W. Liniger:
Exact analysis of an interacting Bose gas: I.
The general solution and the ground state}
{\PR}{130}{(1963), 1605-1616}
\jnlitem{--- :
Exact analysis of an interacting Bose gas: II.
The excitation spectrum}
{\PR}{130}{(1963), 1616-1624}

\refis{L67} \jnlitem
{E. H. Lieb: Residual entropy of square ice}
{\PR}{162}{(1967), 162-172}
\jnlitem{--- :
Exact solution of the F model of an antiferroelectric}
{\PRL}{18}{(1967), 1046-1048}
\jnlitem{---~:~Exact solution of
the two dimensional Slater KDP model of a ferroelectric}
{\PRL}{19}{(1967), 108-110}
\jnlitem{B. Sutherland: Exact solution of
a two dimensional model for Hydrogen bonded crystals}
{\PRL}{19}{(1967), 103-104}

\refis{MM90} \jnlitem
{K.C. Misra and T. Miwa:
Crystal base for the basic representation of $U_{q}(\hat sl(n))$}
{\CMP}{134}{(1990), 79-88}

\refis{Ons44} \jnlitem
{L. Onsager:
Crystal statistics. I.
A two-dimens\-ional model with an order-disorder transition}
{\PR}{65}{(1944), 117-149}

\refis{SW83} \jnlitem
{K. Sogo and M. Wadati:
Boost operator and its application to quantum Gelfand-Levitan
equation for Heisenberg-Ising chain with spin one-half}
{\goodbreak\PTP}{69}{(1983), 431-450}

\refis{Th81} \jnlitem
{H. B. Thacker:
Exact integrability in quantum field theory and statistical systems}
{\RMP}{53}{(1981), 253-285}

\refis{Th86} \jnlitem
{H. B. Thacker:
Corner transfer matrices and Lorentz invariance on a lattice}
{\PD}{18}{(1986), 348-359}

\refis{YY66} \jnlitem
{C. N. Yang and C. P. Yang:
One-dimensional chain of anisotropic spin-spin interactions. I.
Proof of Bethe's hypothesis for ground state in a finite system}
{\PR}{150}{(1966), 321-327}
\jnlitem
{--- :
One-dimensional chain of anisotropic spin-spin interactions. II.
Properties of ground state energy per site for an infinite system}
{\PR}{150}{(1966), 327-338}

\listreferences

\bye